\documentclass[reprint,amsfonts, amssymb, amsmath,  showkeys,aps, superscriptaddress, twocolumn,longbibliography,nofootinbib]{revtex4-2}

\usepackage{graphicx} 

\usepackage[dvipsnames]{xcolor}

\usepackage{float}
\usepackage[shortlabels]{enumitem}

\usepackage{braket}
\usepackage{amsthm}
\usepackage{mathtools}
\usepackage{url}
\usepackage{physics}
\usepackage{graphicx}
\usepackage[left=16mm,right=16mm,top=35mm,columnsep=15pt]{geometry} 
\usepackage[T1]{fontenc}
\usepackage{bm}
\usepackage{tabularx}

\def\HC{\mathcal{H}}

\def\LC{\mathcal{L}}

\def\ad{^{\dagger}}

\def\a{\alpha}

\newcommand{\fsnull}[1]{}
\newcommand{\old}[1]{}

\usepackage[makeroom]{cancel}
\usepackage[toc,page]{appendix}
\definecolor{C1}{RGB}{52, 89, 149}
\definecolor{C2}{RGB}{251, 77, 61}
\definecolor{C3}{RGB}{3, 206, 164}
\definecolor{C4}{RGB}{202, 21, 81}
\usepackage{hyperref}
\hypersetup{colorlinks=true, linkcolor=C4, citecolor=C4, urlcolor=C4}

\usepackage{tikz}
\tikzset{every picture/.style=remember picture}

\usepackage{pgfplots}

\usepackage[utf8]{inputenc}
\usepackage{graphicx}
\usepackage{xcolor}
\usepackage{amsmath}
\usepackage{amsthm}
\usepackage{bm}
\usepackage{bbm}
\usepackage{comment}
\usepackage{appendix}
\usepackage{mathdots}
\usepackage{lipsum}
\usepackage{verbatim}
\usepackage{nccmath}
\usepackage{amsfonts}
\usepackage{thm-restate}
\usepackage{thmtools}
\usepackage{capt-of}

\newtheorem{crl}{Corollary}

\newtheorem{fact}{Fact}

\newtheorem{algorithm}{Algorithm}
\newtheorem{prop}{Proposition}

\usepackage{amssymb}
\usepackage{dsfont}

\newcommand{\poly}{\operatorname{poly}}

\newcommand{\BC}{\mathcal{B}}

\newcommand{\OC}{\mathcal{O}}
\newcommand{\PC}{\mathcal{P}}

\newcommand{\SC}{\mathcal{S}}

\renewcommand{\geq}{\geqslant}
\renewcommand{\leq}{\leqslant}

\newcommand{\spn}{{\rm span}}

\newcommand{\ot}{\otimes}
\newcommand{\ts}{^{\otimes 2}}

\newcommand{\bs}{\textsf{BS}}

\renewcommand{\th}{\theta }

\newcommand{\lm}{\lambda }

\newcommand{\Lm}{\Lambda }

\newcommand{\Om}{\Omega }

\DeclareMathOperator*{\expect}{\mathbb{E}}

\newcommand{\mcn}{\mathcal{N}}
\newcommand{\mcl}{\mathcal{L}}
\newcommand{\mcs}{\mathcal{S}}

\newcommand{\mco}{\mathcal{O}}

\newcommand{\mcg}{\mathcal{G}}
\newcommand{\mcv}{\mathcal{V}}
\newcommand{\mch}{\mathcal{H}}

\newcommand{\mcx}{\mathcal{X}}
\newcommand{\mcp}{\mathcal{P}}

\newcommand{\mce}{\mathcal{E}}
\newcommand{\mcf}{\mathcal{F}}

\newcommand{\mbsp}{\mathbb{SP}}

\newcommand{\mbo}{\mathbb{O}}
\newcommand{\mbu}{\mathbb{U}}
\newcommand{\mbc}{\mathbb{C}}
\newcommand{\mbr}{\mathbb{R}}

\newcommand{\mst}{\mathsf{T}}

\newcommand{\mss}{\mathsf{S}}

\newcommand{\mfg}{\mathfrak{g}}

\def\be{\begin{equation}}
\def\ee{\end{equation}}
\def\bs{\begin{split}}
\def\e{\end{split}}
\def\ba{\begin{eqnarray}}
\def\bea{\begin{eqnarray}}

\def\tea{\end{eqnarray}}
\def\ea{\end{eqnarray}}
\def\eea{\end{eqnarray}}

\def\tkt{^{\otimes k/2}}

\def\a{\alpha}
\def\b{\beta}

\def\a{\alpha}

\def\b{\beta}

\def\g{\mathfrak{g}}

\def\tn{^\otimes n}
\def\tk{^\otimes k}

\def\a{\alpha}
\def\b{\beta}

\newtheorem{lemma}{Lemma}

\def\td{^{\otimes 2}}

\def\tn{^{\otimes n}}

\def\tk{^{\otimes k}}

\newcommand{\id}{\mathds{1}}

\renewcommand{\a}{\alpha}
\renewcommand{\b}{\beta}

\newcommand{\eps}{\epsilon}

\renewcommand{\th}{\theta}

\newcommand{\arr}{\xrightarrow[]{}}

\newcommand{\sdbraket}[2]{ \langle\!\langle#1 | #2 \rangle\!\rangle}

\def\triv{{\rm triv}}

\def\g{\gamma}

\def\triv{{\rm triv}}

\def\poly{{\rm poly}}

\newcommand{\mc}[1]{\mathcal{#1}}

\newcommand\mbb[1]{\mathbb{#1}}

\newtheorem{definition}{Definition}

\usepackage{xr}
\usepackage{zref-xr}

\def\be{\begin{equation}}
\def\te{\end{equation}}
\def\ee{\end{equation}}
\def\ba{\begin{eqnarray}}
\def\bea{\begin{eqnarray}}

\def\tea{\end{eqnarray}}
\def\ea{\end{eqnarray}}
\def\eea{\end{eqnarray}}

\newtheorem{theorem}{Theorem}

\begin{document}

\title{No-go theorems for sublinear-depth group designs}

\author{Maxwell West}
\thanks{westm@lanl.gov}
\affiliation{Theoretical Division, Los Alamos National Laboratory, Los Alamos, New Mexico 87545, USA}
\affiliation{School of Physics, University of Melbourne, Parkville, VIC 3010, Australia}

\author{Diego Garc\'ia-Mart\'in}
\affiliation{Information Sciences, Los Alamos National Laboratory, Los Alamos, New Mexico 87545, USA}

\author{N. L. Diaz}
\affiliation{Information Sciences, Los Alamos National Laboratory, Los Alamos, New Mexico 87545, USA}
\affiliation{Center for Nonlinear Studies, Los Alamos National Laboratory, Los Alamos, New Mexico 87545, USA}

\author{M. Cerezo}
\affiliation{Information Sciences, Los Alamos National Laboratory, Los Alamos, New Mexico 87545, USA}

\author{Martin Larocca}
\thanks{larocca@lanl.gov}
\affiliation{Theoretical Division, Los Alamos National Laboratory, Los Alamos, New Mexico 87545, USA}
\affiliation{Quantum Science Center, Oak Ridge, TN 37931, USA}

\begin{abstract}
Constructing ensembles of circuits which efficiently approximate the Haar measure over various groups is a long-standing and fundamental problem in quantum information theory. Recently it was shown that one can obtain approximate designs over the unitary group with depths scaling logarithmically in the number of qubits, 
but that no sublinear-depth approximate designs exist over the orthogonal group. Here we derive, for groups $G$ possessing an invariant state $G\tk \ket{\Psi}= \ket{\Psi}$ satisfying certain conditions on its entanglement structure, a lower bound on the diamond distance between the $k$\textsuperscript{th} moment operator of any ensemble of elements of $G$, and that of the Haar measure over $G$. We then use this bound to prove that for many groups of interest, no subset of $G$ consisting of sublinear-depth one-dimensional circuits with constant-locality gates can form an approximate $k$-design over $G$. More generally, on a $D$-dimensional lattice, our results imply that such group designs require depths scaling at least as  $n^{1/D}$. Moreover, for most of the groups we consider we find that such ensembles can, with high probability, be distinguished from $k$-designs by a single shot of a constant-depth measurement. Among other examples, we show that there is a \textit{constant} separation between (a) the maximum depth and gate count for which no circuit can approximate even the second moment of random matchgate circuits, and (b) the depth and gate count required to implement the matchgate Haar distribution exactly. We furthermore rule out the existence of sublinear-depth $8$-designs over the Clifford group. Finally, we relax the assumption of working with local gates, and prove the impossibility of obtaining approximate designs over $G$ using \textit{any} circuit comprised of a sublinear number of gates generated by Pauli strings.
\end{abstract}

\maketitle

\section{Introduction}
Random unitaries are ubiquitous in quantum physics and computation, underlying both innumerable applications of quantum computers~\cite{huang2020predicting,knill2008randomized,elben2022randomized,zhao2021fermionic,wan2022matchgate,west2024real,west2026classical,arute2019quantum,king2024triply,van2022hardware} and our understanding of topics as seemingly disparate as quantum chaos~\cite{fisher2023random,nahum2017quantum,nahum2018operator} and black hole physics~\cite{hayden2007black,sekino2008fast}. Unfortunately, constructing circuits which uniformly sample random unitaries -- drawing from the  \textit{Haar measure} on the unitary group~\cite{mele2023introduction} -- requires a number of elementary operations that grows exponentially with the system size. Therefore, Haar random unitaries cannot be implemented on even the relatively modestly-sized quantum computers available today.
Fortunately, in many  applications one requires only the weaker notion of approximately (to some precision $\varepsilon$) matching the first $k$ moments of the Haar measure for some relatively small value of $k$; an ensemble that does so is termed an  \textit{$\varepsilon$-approximate unitary $k$-design}~\cite{mele2023introduction}. Naturally, a significant amount of effort has been devoted to efficiently constructing approximate unitary designs~\cite{harrow2009random,brandao2016local,hunter2019unitary,liu2022estimating,ho2022exact,chen2024efficient,metger2024simple,haah2024efficient,chen2024incompressibility}. These efforts have  culminated in the findings  of Refs.~\cite{schuster2024random,laracuente2024approximate}, where it was shown that for a one-dimensional $n$-qubit system, $\varepsilon$-approximate unitary $k$-designs  can be constructed with a depth of
$L\in\mco\left(\log(n/\varepsilon)\cdot k\hspace{1.mm}\poly\log(k)\right)$, 
and that this scaling is furthermore \textit{optimal} with respect to qubit count.

In many cases~\cite{zhao2021fermionic,wan2022matchgate,low2022classical,goh2023lie,miller2025simulation,hashagen2018real,west2024real,marvian2022restrictions,hearth2025unitary,west2026classical}, however, one is interested in dynamics that belong to a strict subgroup $G$ of the unitary group, for which the following question naturally arises: \textit{Are there subsets of $G$ consisting of sublinear-depth circuits that form approximate $k$-designs over $G$?} 
An argument ruling out the existence of such subsets was sketched in Ref.~\cite{schuster2024random} for the cases of classical reversible circuits, and for the orthogonal group. In this work, we formalize this argument and extend it to groups possessing a (projectively) invariant state; that is, a state $\ket{\Psi}$ satisfying that for all $U\in G,\ U\tk\ket\Psi=e^{i\th_U}\ket\Psi$. Such groups encompass many cases of interest, including matchgate~\cite{knill2001fermionic,jozsa2008matchgates}, Clifford~\cite{bittel2025complete}, orthogonal~\cite{hashagen2018real}, symplectic~\cite{garcia2024architectures} and mixed-unitary circuits~\cite{grinko2022linear,grinko2023gelfand,nguyen2023mixed}.
As a prominent example, we find that in order to form a matchgate $k$-design for any $k\geq 2$, any ensemble of matchgate circuits requires linear depth and quadratic gate count. Given that one can exactly parametrize the full matchgate Haar measure with circuits of depth $3n$ and $n(2n-1)$ gates~\cite{braccia2025optimal}, our results imply constant separations between depths and gate counts which fail to reproduce even the second moment of the uniform matchgate distribution, and those required to implement it \textit{exactly}. The tightness of these bounds also enables us to conclude that the spectral gap $\Delta$ of the second moment operator of any \textit{matchgate 2-expander}~\cite{mele2023introduction} consisting of unitaries which can be compiled into a product of a constant number of elementary matchgate unitaries satisfies $\Delta\in\mco(1/n)$. This can be thought of as a restriction on the speed with which matchgate circuits can scramble, with large (i.e. $\mco(1)$) gaps leading to faster scrambling.   

\begin{figure}[t!]
    \centering
    \includegraphics[width=.75\linewidth]{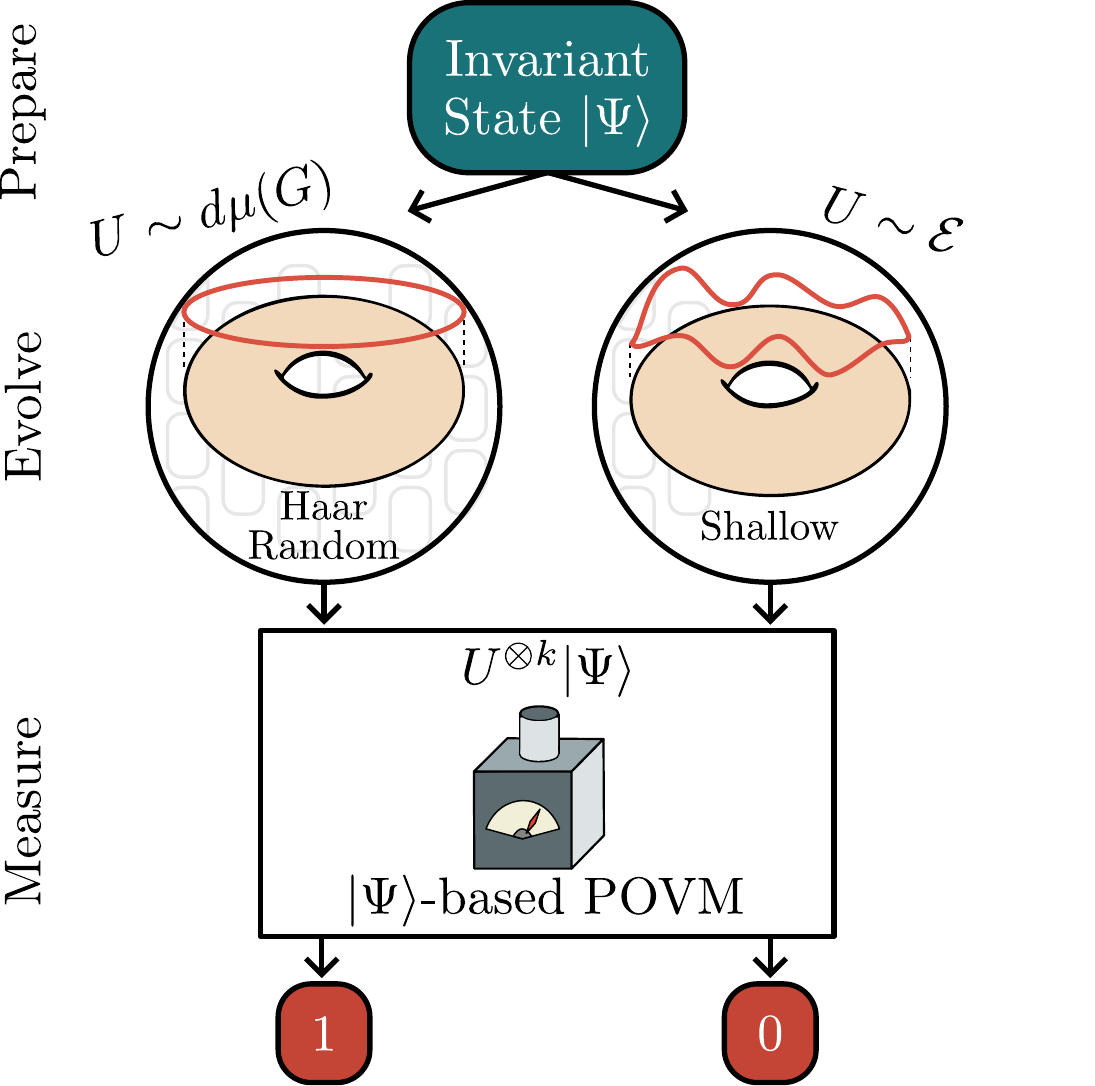}
    \caption{Schematic overview of our main result. The existence of a $G$-invariant state in $\mch\tk$ allows us to construct a POVM that distinguishes between ensembles consisting of shallow local circuits over $G$ and the Haar measure on $G$. In turn, this implies that such ensembles  cannot form $k$-designs over $G$. }
    \label{fig:POVM}
\end{figure}

More generally, our first main result is a simple lightcone-based argument for ruling out group $k$-designs consisting of sublinear-depth one-dimensional circuits composed of constant-support gates from the group (see Fig.~\ref{fig:POVM}). This result has immediate consequences for applications that require approximate group designs: For example, it implies that classical shadow protocols over such groups (many examples of which have recently been proposed~\cite{west2026classical,zhao2021fermionic,wan2022matchgate,west2026particle,heyraud2024unified,west2024real,christensen2026learning,low2022classical,west2024random}) incur an exponential overhead in circuit depth relative to their counterparts over the unitary group. In general, if the individual gates in the circuit have support $\mco(f(n))$, we find that the depth of designs is $\Omega(n/f(n))$.

In our second main result, we go beyond the assumption of local gates and prove no-go results for circuits over ``Pauli compatible'' groups -- groups that are both generated by and have symmetries spanned by Pauli strings.
Specifically, we rule out 2-designs given by ensembles of circuits consisting of a sublinear number of unitaries generated by such not-necessarily-local Pauli generators.
The presence of non-local gates immediately breaks the standard lightcone-based arguments employed in the local case, which we instead generalize to lightcones defined on certain operator-space group modules. Indeed, we will see that these modules come equipped with a natural geometry that allows us to re-introduce a notion of locality, and therefore lightcones.  

Although our results are primarily framed within the language of designs, they may equally be thought of as revealing significant constraints on the ability of many-body quantum systems to scramble information in the presence of  certain symmetries. Indeed, in the presence of these symmetries our results reveal the existence of certain ``probes'' whose characteristics cannot be scrambled away by the system in sublinear time; our bounds operationalize this through the out-of-time-order correlator (OTOC)~\cite{Maldacena_Shenker_Stanford_2016,swingle2018unscrambling,dowling2023scrambling}.

\section{Preliminaries}
Let $\HC \coloneqq (\mbb{C}^2)\tn$ be the state space of an $n$-qubit system on which a compact group of unitaries $G\subseteq \mbu(\mch)$ acts, and denote by $\mcl(\mch)$ the space of linear operators on $\mch$. Our goal is to study the extent to which   certain ensembles of circuits with gates in $G$, with either limited depth or limited gate count, can approximate the uniform measure on $G$. To that end, let us begin by recalling that any ensemble $\mc{E}$ of unitaries on $\HC$ defines (for any $k\geq 1$) the $k$-th moment quantum channel $\phi_{\mc{E}}^{(k)} := \mbb{E}_{U\sim \mc{E}}\Big[ U^{\otimes k}(\cdot) (U\ad)^{\otimes k} \Big]$ on $\LC(\HC^{\otimes k})$
corresponding to the average  action of $U\sim \mc{E}$ on $k$ copies of $\HC$. With this, we can define the notion of approximate designs.
\begin{definition}[Additive $\varepsilon$-approximate $k$-design]
We say that an ensemble $\mc{E}$ is an additive $\varepsilon$-approximate  $k$-design over $G$ if
\begin{equation}\label{eq:dnd}
    \|\phi_{\mc{E}}^{(k)} - \phi_{G}^{(k)}\|_\diamond \leq \eps\,,
\end{equation}
with $\phi_{\mc{E}}^{(k)}$ the average channel for $\mc{E}$ and $\phi_{G}^{(k)}:= \mbb{E}_{U\sim \mu(G)}\Big[ U^{\otimes k}(\cdot) (U\ad)^{\otimes k} \Big]$ the average channel over the Haar measure $d\mu$ on $G$, respectively, and $\|\cdot\|_{\diamond}$  the diamond norm.
\end{definition}
We recall that the diamond distance between a pair of channels $\phi$ and $\phi'$ is given by 
\begin{equation}
\|\phi-\phi'\|_\diamond = \sup_{\rho} \|((\phi-\phi')\otimes\mc{I}_{\rm anc})\rho\|_1, \nonumber
\end{equation}
where the supremum is taken over all  quantum states $\rho$ on $\HC\tk\otimes \HC_{\rm anc}$, with $\HC_{\rm anc}$ an arbitrarily-large ancilla and $\mc{I}_{\rm anc}$ the identity channel on $\LC(\HC_{\rm anc})$.
We note that the strongest notion of a design is measured via the \textit{relative} error rather than the additive error~\cite{schuster2024random}; we here focus solely on the latter since any ensemble that fails to be an additive design will also fail to be a relative design.

Our primary task will be lower-bounding the left hand side of Eq.~\eqref{eq:dnd} for ensembles of sublinear-depth one-dimensional circuits belonging to a wide class of groups. We will do this by leveraging the operational meaning of the diamond norm within the context of channel discrimination that we now describe~\cite{nielsen2000quantum}. 
Given two ensembles $\mc{E}$ and $\mc{F}$ of unitaries over $\HC$, consider the following task:  (i) Prepare a state $\rho \in \LC(\HC^{\otimes k}\otimes \HC_{\rm anc})$, with $\HC_{\rm anc}$ some arbitrarily large ancillary register, (ii) Run a black-box that implements $U^{\otimes k}(\,\cdot\,)(U\ad)^{\otimes k}$ on the system register for $U\sim \mc{E}$ with probability $\lm$ and for $U\sim \mc{F}$ with probability $1-\lm$,  and (iii) Implement a two-outcome positive operator-valued measure (POVM) $\Pi=\{ \Pi_{\mc{E}}, \Pi_{\mc{F}}\}$ over the combined system plus ancillas. If one obtains the measurement outcome $\Pi_\mcx$ (with $\mcx\in\{\mce,\mcf\}$) then one guesses that the applied unitary belongs to the ensemble $\mcx$; the task is to maximize the probability 
\begin{equation}
p_{\rm succ} (\rho,\Pi) = \lm\, p(\Pi_\mc{E} \hspace{0.7mm}|\hspace{0.7mm}U\sim\mce) + (1-\lm)\, p(\Pi_\mc{F}\hspace{0.7mm}|\hspace{0.7mm}U\sim\mcf)\, \nonumber
\end{equation}
of guessing correctly. Importantly,  for any 
choices of $\rho$ and $\Pi$,  
there is a bound on the diamond distance between the channels given by~\cite{benenti_computing_2010} (setting $\lm=1/2$)
\begin{equation}\label{eq:succbound}
    \|\phi_{\mc{E}}-\phi_{\mc{F}}\|_{\diamond} \geq 4(p_{\rm succ}(\rho,\Pi)-1/2)\,.
\end{equation}
Our strategy to derive no-go theorems on approximate group $k$-designs can then be summarized as follows: 
Find a suitable $\rho$ and $\Pi$, compute $p_{\rm succ}(\rho,\Pi)$, and use it to bound the diamond distance between the $k$-th moment operators corresponding to the ensemble of interest and the Haar ensemble, thereby showing that such an ensemble cannot be a $k$-design.

\section{Results}
\begin{table*}\label{tab:1}
\renewcommand{\arraystretch}{1.2}
\begin{tabular}{|c|c|c|c|}
\hline
 \textbf{Group}  & \textbf{Invariant state} & \textbf{No-go result} (this work) & \textbf{Known construction} \\ \hline
   Matchgate    &   $k=2$      &    $L\geq n/2$     &   Haar measure achievable at $L=3n$~\cite{braccia2025optimal}     \\\hline
   Clifford   &    $k=8$     &   $L\geq n-1$       &     Haar measure achievable at $L\in\mco(n)$~\cite{maslov2007linear}         \\ \hline
   Orthogonal   &    $k=2$     &    $L\geq n-1$     &  Orthogonal Cliffords, $L\in\mco(n)$~\cite{hashagen2018real}      \\\hline
   Symplectic   &    $k=2$     &    $L\geq n-1$       & Symplectic Cliffords, $L\in\mco(n)$~\cite{grevink2025will}       \\ \hline
   Mixed-unitary   &    $k=1$     &    $L\geq n-1$       &    Mixed Cliffords, $L\in \OC(n)$ (see App.~\ref{sec:exampledetails})       \\ \hline
   Pauli-compatible   &   Group-dependent      &   Corollary~\ref{crl:pauli}         &      Group-dependent       \\ \hline
\end{tabular}
\caption{An overview of the depth-based restrictions on constructing approximate $k$-designs over various groups proved in this work, with a comparison to previously known results. In this table we assume that the ensembles are over circuits that are constructed from 2-local gates on a one-dimensional geometry (the more general case is considered in the Discussion section). Here $L$ denotes the circuit depth.  }
\end{table*}

We now provide two general constructions, one aimed at bounding circuit depth (Sec.~\ref{sec_cd}), and another aimed at bounding gate count (Sec.~\ref{sec_gc}) of any $k$-design over $G$. 
\subsection{Bounding circuit depth}\label{sec_cd}
Let $\mc{E}_L$ be any ensemble of depth-$L$ circuits composed of gates from $G\subseteq \mbu(\HC)$, and $\mc{E}_G$ be the Haar ensemble on $G$,  with $\phi_L^{(k)}$ and $\phi_G^{(k)}$ the corresponding $k$\textsuperscript{th} moment channels. As previously discussed, we assume that there exists a ``projectively $G$-invariant'' state\footnote{We remark that the existence of a (projectively) invariant state  is equivalent to  $G$ possessing a \textit{preserved bilinear form}; that is, there exists $\Omega\in\mcl(\mch\tkt)$ such that, for all $U\in G,\ U\tkt\Om (U\tkt)^\mst= \chi(U)\Om$ for some character $\chi$ of $G$ (see Appendix~\ref{sec:forms}). Indeed, one simply takes $\ket{\Psi}=(\id\otimes\Omega)\ket{\Phi}$, with $\ket{\Phi}=d^{-k/4}\sum_{i}\ket{ii}$ the Bell state between the two copies of $\HC\tkt$.} $\ket\Psi\in\mch\tk$, i.e., such that for all $U\in G$, $U^{\otimes k} \ket{\Psi} = e^{i\th_U}\ket{\Psi}$  (with $\th_U\in\mathbb{R}$). For technical reasons we will take $k$ to be even. 
Let $V\in \mbu(\HC)$ be a unitary, and consider a bipartition  $\mch  =\mch_L\otimes\mch_{\overline{L}}$, with $\mch_L$ containing the support of $UVU^\dagger$, for all $U\sim\mce_L$. We set $d\coloneqq \dim(\HC),\ $  $d_{\overline{L}}\coloneqq \dim(\HC_{\overline{L}})$. At this stage we do not explicitly make any assumptions about the locality of $V$, but note that failing to choose $V$ to be local may lead to vacuous bounds.  
Our first result is:

\begin{restatable}{thm}{thmmain} \label{thm:main}
If there is a projectively $G$-invariant state in $\mch\tk$ of full Schmidt rank across $\mch\tkt\ot\mch\tkt$, then 
 \begin{equation}\label{eq:mbound}
\|\phi_{\mce_L}^{(k)}-\phi_{\mce_G}^{(k)}\|_\diamond\geq 2-2\expect_{U\sim \mce_G}  \overline{F}_{\mch_{\overline{L}}} (UVU\ad)\,,
\end{equation}
where $\overline{F}_{\mch_{\overline{L}}}(O)=d_{\overline{L}}^{-2}\sum_{P\,:\,{\rm supp}(P)\subset  \mch_{\overline{L}}} d^{-1}   \tr[O\ad P OP ] $ is the  infinite-temperature OTOC averaged over the Pauli probes $P$ with support on $\mch_{\overline{L}}$.  
\end{restatable}
The technical condition on the Schmidt rank is explored further in Section~\ref{sec:dis} and   Appendices~\ref{sec:forms} and~\ref{sec:proofs}; at this point we simply note that it holds in all of our examples.
The proof of Theorem~\ref{thm:main}, which we present in Appendix~\ref{sec:proofs}, is a straightforward application of the previous discussion: Namely, we use the existence of $\ket\Psi$ to construct an experiment that distinguishes between $\mce_L$ and $\mce_G$, and then use Eq.~\eqref{eq:succbound} to translate this into a bound on the difference of the  diamond norm of the moment operators.  
Intuitively, considering an invariant $\ket\Psi\in\mch\ts$, and a local $V\in\mbu(\mch)$. Then the invariance of $\ket\Psi$ implies that $U\ts (V\ot\id)\ket\Psi = (UVU\ad \ot \id)\ket\Psi$, which allows us to make lightcone arguments about the Heisenberg-evolved $V$. 

Note that there is no explicit dependence on $\ket\Psi$; indeed, in principle one does not even need to know what $\ket\Psi$ is, only that it exists\footnote{See Appendix~\ref{sec:rep} for a representation-theoretic discussion of how one could non-constructively deduce the existence of such a state.}.
All that is then left is to evaluate the Haar average on the RHS of Eq.~\eqref{eq:mbound} for various choices of $G$ and $V$. Intuitively, if a random unitary from $G$ typically scrambles $V$ into something which acts  non-trivially on $\mch_{\overline{L}}$, then the bound of Eq.~\eqref{eq:mbound} is strong. For example, if $V$ is twirled by a typical $U$ into an approximately uniform sum over all Pauli strings, then we will have $\expect_{U\sim \mce_G} \overline{F}_{\mch_{\overline{L}}} (UVU\ad)\sim (\dim \mch_{\overline{L}})^{-2}$, which we can easily take to be much less than one. Indeed, (with the exception of the matchgate group), this is exactly the scaling that we observe in our examples.
In the following, we leverage the previous tools to prove several no-go results for sublinear-depth approximate designs. In all of our examples we will assume a one-dimensional connectivity, but we emphasize that Theorem~\ref{thm:main}  applies to any geometry, a point to which we will return in the Discussion section. 

\subsubsection{Matchgate group}
For our first example, we consider the $n$-qubit matchgate (or free-fermion) group ${\rm Spin}(2n)$, which we recall to be generated by nearest-neighbor (on a one-dimensional array) $XX$ gates and single-qubit $Z$ gates~\cite{diaz2023showcasing,knill2001fermionic,jozsa2008matchgates}. We show in Appendix~\ref{sec:forms} that a suitable matchgate-invariant state exists for $k=2$. Applying Theorem~\ref{thm:main} with $V=X_{n/2}$ and $L=n/2-1$ 
leads to 
\begin{equation}\label{eq:mgstir}
    \|\phi_{L}^{(2)}-\phi_{{\rm Spin}(2n)}^{(2)}\|_\diamond\geq 2-\frac{n+1}{2n-1}\approx 3/2\,,
\end{equation}
from which we conclude (in Appendix~\ref{sec:exampledetails}) that
\begin{crl}\label{crl:mg1}
Any ensemble of matchgate circuits requires depth $L\geq n/2$ to form an approximate matchgate $2$-design.
\end{crl}
As the Haar measure over the matchgate group can be exactly sampled from with circuits of depth $3n$~\cite{braccia2025optimal}, Corollary~\ref{crl:mg1} demonstrates that there is a constant separation between the depth required to exactly reproduce all moments of the uniform matchgate distribution, and the depth required to merely approximately reproduce its second moment.

\subsubsection{Orthogonal group}
\noindent
For our next example, we consider the orthogonal group,
\begin{equation}
\mbo(d)=\{U\in\mbu(d)\ :\ U^\mst U = \id\}.
\end{equation}
As described in the Introduction, the non-existence of  sublinear-depth approximate orthogonal 2-designs consisting of local orthogonal unitaries has already been sketched in Ref.~\cite{schuster2024random}.
As the Bell state $\ket\Phi=2^{-n/2}\sum_i\ket{ii}$ is an orthogonal-invariant state in $\mch\ts$, Theorem~\ref{thm:main} applies, and we turn to the evaluation of Eq.~\eqref{eq:mbound}, with $V=Z_1$, the Pauli-$Z$ operator on the first qubit. As we show in Appendix~\ref{sec:exampledetails}, any depth-$L$ ensemble $\mce_L$ of orthogonal unitaries built from 2-local gates satisfies 
\begin{equation}\label{eq:obound}
\|\phi_L^{(2)}-\phi_{\mbb{O}(d)}^{(2)}\|_\diamond\geq 2\left(1-\frac{d_L^2 +d_L-2 }{(d+2)(d-1)}\right)\,.
\end{equation}
Taking for example $L=n-2$, and correspondingly $\mch_L$ to be the first   $n-1$ qubits, one sees that $d_L=d/2$ and that therefore, any ensemble with $L=n-2$ satisfies $\|\phi_{n-2}^{(2)}-\phi_{\mbo(d)}^{(2)}\|_\diamond\gtrsim 3/2
$, whence
\begin{crl}\label{crl:ortho}
Any ensemble of circuits consisting of $2$-local orthogonal gates requires depth $L\geq n-1$ to form approximate orthogonal $2$-designs.
\end{crl}
As real-valued Cliffords form orthogonal 2-designs and can be implemented in linear depth~\cite{hashagen2018real}, it follows that amongst subsets of the orthogonal group they are, up to a constant factor, depth-optimal for this purpose.

\subsubsection{Unitary symplectic group}
The situation is very similar for the (unitary) symplectic group, which we recall to be defined by the condition 
\begin{equation}
\mbsp(d/2)=\{U\in\mbu(d)\ :\ U^\mst\Om U = \Om\}\,,
\end{equation}
for a fixed antisymmetric non-degenerate bilinear form which we will take to be $iY_1$\footnote{This choice  identifies a specific instantiation of the symplectic group; a different choice (necessarily related by an orthogonal change of basis) would lead to an isomorphic copy.}~\cite{garcia2024architectures,west2024random}. Here, the state $(\id\otimes\Omega)\ket{\Phi}$, where $\ket{\Phi}$ is the Bell state, is a $G$-invariant state in $\HC^{\otimes 2}$.
Thus, we are once again in a position to apply Theorem~\ref{thm:main}. 
Taking $V=Z_1$, we show in Appendix~\ref{sec:exampledetails} that any depth-$L$ ensemble $\mce_L$ of symplectic unitaries built from 2-local gates satisfies
\begin{equation}
\|\phi_L^{(2)}-\phi_{\mbsp(d/2)}^{(2)}\|_\diamond \geq 2\left(1-\frac{ d_L(d_L+1)}{d(d+1)}\right)\,,
\end{equation}
and, taking $L=n-2$ and $\mch_L$ to be the first $n-1$ qubits we find $\|\phi_{n-2}^{(2)}-\phi_{\mbsp(d/2)}^{(2)}\|_\diamond \gtrsim 3/2$, from which we conclude
\begin{crl}\label{crl:symp}
Any ensemble of $2$-local symplectic gates requires depth $L\geq n-1$ to form an approximate symplectic $2$-design.
\end{crl}
\vspace{-7mm}

\subsubsection{Clifford  group}
We next come to the important case of the $n$-qubit Clifford group~\cite{bittel2025complete}. This is our first example of a group for which there is no invariant state at $k=2$. Indeed, the Clifford group is well-known to form a unitary 3-design~\cite{mele2023introduction}, so the existence of logarithmic-depth approximate unitary designs immediately implies the existence of logarithmic-depth approximate Clifford 3-designs. As  Clifford unitaries do not form a unitary 4-design~\cite{zhu2016clifford}, however, the question of the existence of sublinear-depth Clifford designs for $k>3$ is open. In Appendix~\ref{sec:exampledetails} we demonstrate the existence of a Clifford-invariant state at $k=8$, and then use a slight modification of Theorem~\ref{thm:main} to find the following result:
\begin{crl}
Any ensemble built from $2$-local Clifford gates requires depth $L\geq n-1$ to form an approximate Clifford $8$-design.
\end{crl}
As the Clifford unitaries themselves can be implemented in linear depth~\cite{maslov2007linear}, one can certainly construct linear-depth Clifford $8$-designs. This means  that our bound  is again tight up to constant factors. 

\subsubsection{Mixed-unitary group}
We next study \textit{mixed-unitary} circuits~\cite{grinko2023gelfand}, which consist of elements of the form $U\ot U^*$, where $U$ is an $n$-qubit unitary. Here for the first time we have an invariant state at $k=1$. Indeed, note that the Bell state $\ket{\Phi}$ across the first and second registers of $n$ qubits is manifestly invariant. By a  routine application of our procedure we find 
\begin{crl}\label{crl:mixedu}
Any ensemble built from $4$-local mixed-unitary circuits (i.e., $U\ot U^*$ with $2$-local $U$) requires depth $L\geq n -1$ to form an approximate mixed-unitary $1$-design.
\end{crl}
We show in Appendix~\ref{sec:exampledetails} that the mixed Clifford ensemble ${\rm Cl}_n\ot {\rm Cl}_n^*$ is a mixed-unitary 1-design, so that our bound is optimal to a constant factor\footnote{Note, for example, that although the Pauli group $\mcp_n$ forms a (constant-depth) unitary 1-design, $\mcp_n\ot\mcp_n^*$ is \textit{not} a mixed-unitary 1-design (see Appendix~\ref{sec:exampledetails}).}.

\subsubsection{Pauli-compatible groups}
As our last example of Theorem~\ref{thm:main} we consider the  class of \textit{Pauli-compatible} groups, defined as follows.  

\begin{definition}[Pauli-compatible]\label{def:pcg}
Let $\PC_n$ be the set of $n$-qubit Pauli strings.
A group $G\subseteq \mbb{U}(\HC)$ on $\HC=(\mbb{C}^2)\tn$ is called Pauli-compatible if both
\begin{itemize}
    \item its elements are of the form $\prod_j e^{i\theta_j P_j} $ with $\th_j$ real scalars and $P_j \in \SC \subseteq \PC_n$, and
    \item the $G$-invariant operator space has a Pauli basis: $\LC(\HC)^G = \spn_\mbc\, \BC$, with $\BC\subseteq \PC_n$.
\end{itemize}
\end{definition}

This definition is quite general, including for example the matchgate, orthogonal and symplectic groups, as well as the unitary group itself. The benefit of working with Pauli-compatible groups comes from their connection to the next (and final) object we need to define, that of a \textit{commutator graph}~\cite{diaz2023showcasing,west2025graph}:
\begin{definition}[Commutator graph]\label{def:cg}
The commutator graph $\mcg_\mcs$ of a set $\mcs$ of $n$-qubit Pauli strings is the graph with a vertex for every Pauli string (all $4^n$ of them, not just those in $\mcs$), and an edge between the vertices corresponding to the Pauli strings $P$ and $Q$ iff there is a $H\in\mcs$ such that $[H,P]\propto Q$.
\end{definition}
So, for our final application of Theorem~\ref{thm:main}, we consider $G$ a Pauli-compatible group.
As we show in Appendix~\ref{sec:forms},  any such $G$ possesses an invariant state at $k=2$ 
if and only if there is a non-zero  $\Om\in\mcl(\mch)$ such that for every $P\in\mcs$,   $P\Om +\Om P^\mst=0$. 
Leveraging Eq.~\eqref{eq:mbound}, 
we find (see Appendix~\ref{sec:exampledetails} for a proof) that 

\begin{crl}\label{crl:pauli}
Any ensemble over depth-$L$ 2-local circuits from  a Pauli-compatible group $G$ with generators $\mc{S}$ and an invariant form $\Om$ satisfies, for any Pauli string $P$, 
 \begin{equation}\label{eq:boundpauli}
\|\phi_{L}^{(2)}-\phi_{G}^{(2)}\|_\diamond\geq 2(1-r_G(P,L)),
\end{equation}
where $r_G(P,L)$ is the fraction of Pauli strings in the same connected component as $P$ of $\mcg_\mcs$ (recalling Definition~\ref{def:cg})  
   that act non-trivially only within $L$ qubits of the support of $P$. 
\end{crl}
For example, in the matchgate case, we find
\begin{align}
    r_{{\rm Spin}(2n)}(X_{n/2}, n/2-1) &= {2n-2 \choose n-1}{2n \choose n-1}^{-1}\approx 1/4\,.
\end{align}
Note that using $r_{{\rm Spin}(2n)}(X_{n/2}, n/2-1)$ in  Corollary~\ref{crl:pauli} immediately yields Eq.~\eqref{eq:mgstir}. Here the perturbation $V=X_{n/2}$ was chosen due to the fact that $X_{n/2}$ lives in a commutator graph component of size $\Theta(\exp( n))$~\cite{diaz2023showcasing,west2025graph}, which facilitates bounding $r_{{\rm Spin}(2n)}$ away from one (see Appendix~\ref{sec:exampledetails}). 

\subsection{Bounding gate count}\label{sec_gc}

Up to this point we have showcased the use of  Theorem~\ref{thm:main} to rule out sublinear-depth designs over a wide variety of groups of interest, so  long as the circuits are composed of constant-locality elementary gates. While this is hardly an unreasonable assumption,
we now present a way of relaxing it.
Indeed, we can appeal to a generalized notion of locality in the \textit{operator space} $\mcl(\mch)$ to distinguish between ensembles of shallow and deep circuits composed of gates of arbitrary geometric locality and bodyness. In fact, for a Pauli-compatible group of unitaries of the form 
\begin{equation}\label{eq:unit}
U=\prod_{\ell=1}^N\exp(i\th_\ell P_\ell),
\end{equation}
where $P_\ell \in \mcs\ \forall\ell$, 
we can use the commutator graph $\mcg_\mcs$ to define such a generalized notion of locality.
Indeed, consider some Pauli string $P$. For a $U$ of the above form, the Heisenberg-evolved operator $U^\dagger PU$ will spread across (and only across) the connected component of the commutator graph in which $P$ resides~\cite{diaz2023showcasing,west2025graph}. Moreover, and as we show in Appendix~\ref{sec:proofs}, the action of an individual gate $U_\ell = \exp(i\th_l P_l)$ evolves a Pauli only into a linear combination of itself and one of its \textit{nearest neighbors} on the graph\footnote{The key point is that conjugating $Q$ by $e^{i\theta P}$ acts as a rotation by $2\th$ in the plane spanned by $Q$ and $[P,Q]$.}. 
It is in this sense that we recover a notion of locality (this time in operator space) from which we  obtain:
\begin{restatable}{thm}{thmnlocal}  \label{thm:nlocal}
Let $\mce_N^\mcs$ be an ensemble of circuits  of the form $U=\prod_{\ell=1}^N\exp(i\vartheta_\ell H_\ell)$, where  $\{H_\ell\}_\ell\subset\mcs$ are Pauli strings in the Lie algebra of a Pauli-compatible, bilinear-form-preserving group $G$. Then, 
 \begin{equation}\label{eq:thm2}
\|\phi_{\mce_N^\mcs}^{(2)}-\phi_{\mu_{G}}^{(2)}\|_\diamond\geq 2-2\frac{\left\lvert \mcn_{N,\mcs}(P)\right\rvert}{\left\lvert \mcn_{\mcs}(P)\right\rvert},
\end{equation}
where $\mcn_{N,\mcs}(P)$ denotes the set of vertices within a distance of at most $N$ of $P$ in the commutator graph corresponding to the generating set $\mcs$, and $\mcn_{\mcs}(P)$ the connected component of the graph containing $P$.
\end{restatable}
To further clarify the meaning of Theorem~\ref{thm:nlocal}, we  provide the following intuition. A Pauli string $P$ evolved by a circuit $U$ of the form  in Eq.~\eqref{eq:unit} will gradually spread across the corresponding connected component in the commutator graph. If $U$ consists of $N$ gates (letting ``gate'' refer to  an elementary constituent $\exp(i\vartheta_\ell H_\ell)$ of the circuit), $P$ will only propagate throughout its neighborhood $\mcn_{N,\mcs}(P)$. On the other hand, a Haar random circuit  will spread $P$  (with high probability) approximately uniformly across the entire component $\mcn_{\mcs}(P)$~\cite{west2025graph}. If $\lvert \mcn_{N,\mcs}(P)\rvert\ll \lvert \mcn_{\mcs}(P)\rvert$, one would expect to be able to differentiate these behaviors. The proof of Theorem~\ref{thm:nlocal} (which may be found in Appendix~\ref{sec:proofs}) is just the formalization of this intuition, combined as previously with the observation that Eq.~\eqref{eq:succbound} allows us to translate distinguishability into a statement about diamond-norm differences.

It follows from the previous discussion that we can consider a general Pauli-compatible group $G$, making no assumptions on the support of the Pauli strings  in the generating set $\mcs$.  For example,  Eq.~\eqref{eq:thm2} implies that for any such ensemble $\mce_N^\mcs$ comprising circuits with at most $N$ gates, we have for any Pauli $P$ that
\begin{equation}\label{eq:lb}
\|\phi_{\mce_N^\mcs}^{(2)}-\phi_{\mu_{G}}^{(2)}\|_\diamond \geq 2\left(1 - \frac{\left\lvert \mcs\right\rvert^N+1}{\left\lvert \mcn_{\mcs}(P)\right\rvert} \right).
\end{equation}
If the commutator graph of $\mcs$ has a component of size $\Om(\exp(n))$, and when $\abs{\mcs}\in \mco({\rm poly}(n))$, then we immediately see that no ensemble composed of a logarithmic number of gates can well-approximate a 2-design over $G$. 
While it is perhaps already interesting that  one can make statements at this level of generality, we will see that Eq.~\eqref{eq:lb} is in general quite loose, and that directly applying Eq.~\eqref{eq:thm2} to specific generating sets can lead to exponentially tighter bounds.

For example, let us apply Theorem~\ref{thm:nlocal} to  matchgate circuits with one- and two-qubit gates.  We show in Appendix~\ref{sec:exampledetails} that one has
\begin{crl}\label{crl:mgg}
Any ensemble over 2-local matchgate circuits consisting of less than  $n^2/2$ gates can be efficiently distinguished from a matchgate 2-design.
\end{crl}

As one can implement Haar-random matchgates using $2n^2-n$ local gates~\cite{braccia2025optimal} (i.e., the dimension of the corresponding Lie algebra), our results therefore establish a constant separation between gate counts that cannot reproduce even the second moment of Haar-random matchgate circuits, and fully Haar-random matchgates. 

Interestingly, the super-linear nature of the lower bound of Corollary~\ref{crl:mgg}  has non-trivial implications for the spectra of \textit{matchgate $t$-expanders}, where we recall the definition~\cite{mele2023introduction} 
\begin{definition}[Expander over $G$]\label{def:texp}
We call an ensemble $\mce$ of unitaries in $G$ a $\lm$-approximate $t$-expander over $G$ if 
\begin{equation}\label{eq:texp}
    \| \phi^{(t)}_\mce - \phi^{(t)}_{\mu_G}\|_{\infty} \le \lm
\end{equation}
for  $\lm<1$. 
\end{definition}
That is, an expander is an approximate design with respect to the spectral norm. One of the primary sources of interest in the notion of expanders is that they can be \textit{amplified}; indeed if $\nu$ is a $\lm$-approximate $t$-expander over $G$ then the ensemble formed by taking  products of $p$ iid elements from $\nu$ is a $\lm^p$-approximate $t$-expander over $G$~\cite{mele2023introduction}. Evidently the rate at which taking ``multiple steps'' of such a channel will converge to the corresponding uniform channel $\phi^{(t)}_{\mu_G} $ depends on the \textit{spectral gap} $\Delta:=1-\lm$, the calculation of which is therefore of considerable  interest~\cite{harrow2009random,brandao2016local,hunter2019unitary,haferkamp2022random,haferkamp2021improved,chen2024incompressibility,deneris2024exact}. We are able to connect the spectral gap of a 2-expander to the number of elementary gates it employs as follows. Suppose that a single step of an expander consists of $g(n)$ gates, and that we have a bound $\|\phi_{\mce_{N}}^{(2)}-\phi_{\mu_{G}}^{(2)}\|_\diamond\geq c$ of the form of Theorem~\ref{thm:nlocal} for some constant $c>0$. We show in Appendix~\ref{sec:exampledetails} that $\Delta \lesssim \log (4)g(n)n / N$, from which (combined with Corollary~\ref{crl:mgg}) it follows  that
\begin{crl}\label{crl:mgexp}
Any matchgate 2-expander whose one-step channel uses $\mco (1)$ elementary matchgates has $\Delta\in\mco(1/n)$.
\end{crl}
We emphasise that such a (non-constant) bound on the spectral gap was possible only as Corollary~\ref{crl:mgg} allows us to take $N\in\mco(n^2)$.

For the final result of this section, we consider the case of
$N$-gate-circuit ensembles over ``non-local matchgate circuits''; that is, circuits of the form
\begin{equation}
    U=\prod_{\ell=1}^{N}\exp(i\vartheta_\ell H_\ell)\,,
\end{equation}
where each $H_\ell$ can be any Pauli string in the Lie algebra $\mathfrak{so}(2n)$ associated to the Lie group ${\rm Spin}(2n)$. For convenience we recall that
\begin{equation}\label{eq:match-DLA}
\mathfrak{so}(2n) \cong {\rm span}_\mbr i\{Z_i,\widehat{X_iX_j},\widehat{X_iY_j},\widehat{Y_iX_j},\widehat{Y_iY_j}\}_{1\leq i< j\leq n},
\end{equation}
with $\widehat{A_iB_j}:=A_i Z_{i+1}Z_{i+2}\cdots Z_{j-1}B_j$~\cite{diaz2023showcasing}. In this setting, the arbitrarily non-local support  of the allowed generators immediately causes the lightcone-based arguments that  underlie Theorem~\ref{thm:main} to fail. Nonetheless, we show in  Appendix~\ref{sec:exampledetails} that circuits of this form cannot form approximate matchgate 2-designs. For example, taking  $N=n/c$ for any fixed $c>2$ we find
\begin{equation}
\|\phi_{\mce_{N=n/c}}^{(2)}-\phi_{\mu_{{\rm Spin}(2n)}}^{(2)}\|_\diamond\gtrsim  2 - \mco\left(\sqrt{n}b^{-n}\right)\,,
\end{equation}
for some $b>1$. That is, 
\begin{crl}\label{crl:mgarb}
No ensembles over sublinear-gate-count-circuits composed of unitaries generated by arbitrary Pauli strings in the Lie algebra associated to matchgates (Eq.~\eqref{eq:match-DLA}) well-approximate the second moment of the matchgate Haar measure. 
\end{crl}

\section{Discussion}\label{sec:dis}
In this work we have proven that the failure of ensembles of shallow local circuits to match the Haar-average values of certain quantities can be exploited to rule out short-depth approximate designs over groups which possess an invariant state. There is then an intriguing contrast between this result and the fact that approximate designs over the entire unitary group can be achieved in logarithmic  depth~\cite{schuster2024random,laracuente2024approximate}, a task which one might have imagined to be generically harder. 
Certainly sublinear-depth ensembles over the full unitary group also fail to match certain properties of Haar-random unitaries~\cite{schuster2024random}; the key point is that the ways in which they fail cannot be distinguished by any observer with the ability to single-shot sample a $U\tk$ for constant $k$.  Conceptually, the important difference in the cases considered here is that the invariant state allows one to time-reverse copies of the unitaries sampled from the ensemble, a resource that is known to be powerful for learning~\cite{schuster2024random,king2024exponential}. 
Framing the construction of approximate designs as a competition against an adversary who attempts to discern one's proposed ensembles from true designs, then, we see  the source of the additional difficulty of constructing designs over these groups. In a somewhat similar vein, the recent work Ref.~\cite{haah2025short}  proves that the presence of conserved quantities rules out sublinear-depth designs over the group of unitaries that conserve them, complementing our results on the  nonexistence of designs over subgroups of the unitary group with invariant states. 

Beyond their intrinsic interest, the extreme prevalence of designs in quantum computation and information yields a number of immediate practical consequences of our results. For example, in the hugely popular framework of classical shadows~\cite{huang2020predicting}, one typically performs randomized evolution from a 3-design over a certain group; group 3-designs allow one to leverage well-established group integration techniques to compute average shadow channels (and their inverses) and bounds on the variance of the estimators~\cite{huang2020predicting}.
In the case of classical shadows over the unitary group,  one can then employ Cliffords, famously forming a unitary 3-design as they do~\cite{webb2016clifford}. Arbitrary Cliffords, however, require linear depth to be implemented by local operators; a key application of Ref.~\cite{schuster2024random} is to formalize recent work~\cite{bertoni2024shallow} showing that one can drastically reduce this requirement. On the other hand, there has recently been much interest in performing classical shadows over different groups; our results rule out similar improvements for classical shadow protocols defined over the matchgate~\cite{wan2022matchgate,zhao2021fermionic,west2026particle,heyraud2024unified,christensen2026learning,low2022classical}, orthogonal~\cite{west2024real,liang2024real} and symplectic~\cite{west2024random} groups. Similarly, designs over the orthogonal and matchgate groups are needed in the corresponding randomized benchmarking protocols~\cite{hashagen2018real,helsen2022matchgate}.

We reiterate that our results apply only to $G$-designs by ensembles over elements that are themselves elements of $G$. Crucially, it is sometimes possible to form $G$-designs from ensembles that are not subsets of $G$. As a simple example, uniformly sampling Paulis yields an orthogonal 1-design, but not all Paulis are in the orthogonal group. It is therefore an important open question whether our no-go results can be circumvented by considering more general such ensembles; their existence would have immediate consequences for the applications of the previous paragraph.

Next, we remark that in our example applications of Theorem~\ref{thm:main}, we have for simplicity assumed that the ensembles are defined over quantum circuits on a one-dimensional geometry. Theorem~\ref{thm:main} is however geometry-agnostic, holding for any bipartition of a Hilbert space $\mch$ into the tensor product of a factor containing the support of some initial operator, and its tensor complement in $\mch$. For example, if an $n$-qubit circuit is defined on a $D$-dimensional grid, a simple generalization of our results implies designs require depths scaling at least as $\Om(n^{1/D})$ when employing constant-locality gates. In the most general case, for qubits on a connectivity graph of diameter $D$, and for gates of locality $\ell$, for highly scrambling groups we will obtain lower bounds of $\Omega(D/\ell)$. 

Another interesting aspect to note is that in most of our constructions, although not needed for the purposes of ruling out the designs, the channel discrimination experiments consist of efficiently preparable states and measurements, and may thus be of interest in the context of property testing~\cite{montanaro2013survey}; more precisely, 
in testing whether certain experimental ensembles do or do not form  group designs.

Next, let us comment on the origin of the technical requirement in Theorem~\ref{thm:main} of the invariant state being of full Schmidt rank (which is perhaps somewhat counterintuitive, as the Schmidt rank is   not robust to small perturbations). Intuitively, the way that this full Schmidt rank condition appears in our calculations (see Appendices~\ref{sec:forms} and~\ref{sec:proofs}) is that we are able to construct from such an invariant state a Bell state, and then ``wrap around'' unitaries from the group in such a way as to make the corresponding ``time-reversed'' unitaries appear, leading to our light-cone  arguments (with all this  ultimately coming from the $(U\ot \id)\ket\Phi=(\id\ot U^{\mathsf{T}})\ket\Phi$ identity, with $\ket\Phi$ the Bell state).  If the invariant state were rank-deficient, we would be able to time-reverse the unitaries only on some strict subspace of the full Hilbert space, which would  not necessarily enjoy favourable locality properties. So, it is unclear that our arguments would carry over to this more general setting (almost certainly not in complete generality, anyway; possibly one could find  individual examples that work out).

Finally, we  comment on the role of complex numbers in logarithmic-depth designs; as Ref.~\cite{schuster2024random} highlights, complex numbers are necessary. At least when the Hilbert space $\mch$ is irreducible as a (complex) representation of $G$, our results strengthen this requirement in a natural way. Specifically, we prove in Appendix~\ref{sec:rep} that there are no sublinear-depth approximate $G$-designs when $\mch$ is a complexification of a \textit{real} representation (i.e., morally real); not only are complex numbers necessary, they must appear in a non-trivial capacity. Perhaps more surprisingly, we further show that if $\mch$ is \textit{quaternionic}~\cite{fulton1991representation} (i.e., despite being merely a representation over $\mbc$, admits a quaternionic structure), the construction also fails. So, for short-depth approximate designs of groups acting irreducibly, complex numbers are necessary but not sufficient, as both real complexified and quaternionic representations fall short. What is needed are complex \textit{representations}.

\medskip

\textit{Note added.} Shortly after this work was posted to the arXiv, Ref.~\cite{grevink2025will} established a sequence of very similar results.  The reader is encouraged to review that manuscript for a complementary perspective on our findings.

\medskip

\section{Acknowledgments}

We acknowledge useful conversations with Ricard Puig. This work was supported by Laboratory Directed Research and Development (LDRD) program of Los Alamos National Laboratory (LANL) under project number 20230049DR.  MW acknowledges the support of the Australian government research training program scholarship and the IBM Quantum Hub at the University of Melbourne. NLD was supported by the Center for Nonlinear Studies at LANL. DGM, MC and ML acknowledge support from LANL's ASC Beyond Moore’s Law project. M.L. was also supported by the Quantum Science Center (QSC), a national quantum information science research center of the U.S. Department of Energy (DOE).

\bibliography{refs,quantum}

\appendix

\onecolumngrid

\section{Invariant states and preserved bilinear forms}\label{sec:forms}
\noindent
We begin this section by showing
\begin{restatable}{lemma}{lems1}\label{lem:supp1}
    The existence of a projectively $G$-invariant state $\ket\Psi$ in $\mch\ts$ is equivalent to the existence of a  projectively $G$-invariant bilinear form $\Om:\mch\ts\to\mbc$; that is, if $U\ts\ket\Psi=\chi(U)\ket\Psi$ for all $U\in G$ and some character $\chi$ of $G$, then $U\Om U^\mst = \chi(U)\Om$. 
\end{restatable}
\begin{proof}
 Fixing a basis for $\mch$ (say, the computational basis) we can simply think of $\Om$ as a matrix (i.e., with matrix elements $\Om_{i,j}=\Om(\boldsymbol{e}_i,\boldsymbol{e}_j)$ for basis elements $\boldsymbol{e}_i,\boldsymbol{e}_j$); we will henceforth make no distinction between bilinear forms and matrices. So, suppose $\ket\Psi$ is projectively invariant,  and let us expand it in the chosen basis as $\ket\Psi=\sum_{i,j}\Om_{j,i}\ket{ij}$. We then have
\begin{align}
    U\ts\ket\Psi&=U\ts \sum_{i,j}\Om_{j,i}\ket{ij} \nonumber\\
    &=\sum_{i,j}\Om_{j,i}U\ket{i}\ot U\ket{j} \nonumber\\
    &=\sum_{i,j,k,l}\Om_{j,i}U_{k,i}U_{l,j}\ket{kl} \nonumber\\
    &=\sum_{k,l}\left(U \Om U^\mst\right)_{l,k}\ket{kl}\,.
\end{align}
But by assumption $U\ts\ket\Psi=\chi(U)\ket\Psi$; 
comparing coefficients we see that $\chi(U)\Om = U \Om U^\mst$; that is, $\Om$ is a projectively $G$-invariant bilinear form. Conversely, given a projectively $G$-invariant bilinear form $\Om$, we can define $\ket\Psi=(\id\ot\Omega)\ket{\Phi}$, where $\ket\Phi=d^{-1/2}\sum_i\ket{ii}$ is a Bell state across the two copies of $\mch$. We then have:
\begin{equation}
    U\ts\ket\Psi = U\ts (\id\ot\Omega)\ket{\Phi}= (\id\ot (U\Omega U^\mst))\ket{\Phi}=\chi(U) (\id\ot \Omega) \ket{\Phi}=\chi(U)\ket\Psi,
\end{equation}
where we have used the \textit{transpose trick}, $(A\ot B)\ket\Phi =(\id\ot (BA^\mst))\ket\Phi$.
\end{proof}

In our applications we will  require $\Om$ to be \textit{non-degenerate}, a distinction which does not follow automatically from its being a projectively preserved bilinear form of a group action. We do automatically have non-degeneracy in the following important case, however:
\begin{restatable}{lemma}{lems2}
If $\mch$ is irreducible under the action of $G$, then any non-zero projectively $G$-invariant bilinear form $\Om$ is non-degenerate.
\end{restatable}
\begin{proof}
As $\Om$ is non-zero there exists a non-zero $v\in \mathfrak{Im}(\Om)$; that is, there exist non-zero $v,w$ such that $v=\Om w$. We will show that $\mathfrak{Im}(\Om)$ is a representation of $G$. Indeed, for $U\in G$ we have $Uv=U\Om w=\Om(\chi(U)U^*w)\in\mathfrak{Im}(\Om)$, so that $\mathfrak{Im}(\Om)$ is a $G$-invariant subspace of $\mch$. But then by the irreducibility of $\mch$, $\mathfrak{Im} \ \Om=\mch$, and so the kernel of $\Om$ is trivial. That is, $\Om$ is non-degenerate. 
\end{proof}
\noindent
Interestingly, we can use a little trick to promote any invertible projectively $G$-invariant $\Om$ to a \textit{unitary} projectively $G$-invariant $\widetilde \Om$. Indeed,
\begin{restatable}{lemma}{lems3}\label{lem:supp3}
If a  group $G$ possesses a non-degenerate projectively invariant bilinear form $\Om$, then it also possesses a \textit{unitary} projectively invariant bilinear form $\widetilde\Om$.
\end{restatable}
\begin{proof}
Let $P=\Om \Om^\dagger > 0$. We claim that $\widetilde\Om = P^{-1/2}\Om$ is a unitary projectively $G$-invariant bilinear form. The unitarity is immediate: $(\widetilde\Om) (\widetilde\Om)^\dagger=P^{-1/2} \Om\Om^\dagger (P^{-1/2})^\dagger=P^{-1/2} P P^{-1/2}=\id$. To demonstrate the projective $G$-invariance of $\widetilde{\Om}$ we begin by noting that $P$ commutes with $G$; indeed for $U\in G$ we have $UP=U\Om\Om^\dagger=\chi(U)\Om U^*\Om^\dagger=\Om\Om^\dagger U$, where we have used $U\Om U^\mst=\chi(U)\Om $ and the unitarity of $U$. As one commutes with a strictly positive operator $P$ if and only if one commutes with $P^{-1/2}$, we can now rapidly verify the projective invariance of $\widetilde{\Om}$:  
\begin{equation}
U\widetilde{\Om}U^\mst = UP^{-1/2}{\Om} U^\mst = P^{-1/2}U{\Om} U^\mst=\chi(U) P^{-1/2}{\Om}U^* U^\mst= \chi(U)P^{-1/2}{\Om}=\chi(U)\widetilde{\Om},
\end{equation}
concluding the proof.
\end{proof}
While the above discussion was for notational  simplicity phrased in terms of a projectively invariant state in $\mch\ts$, one immediately has the following generalization: 
If one has a projectively invariant state $\ket\Psi\in\mch\tk$ which is of full Schmidt rank across  $\mch^{\otimes k/2}\ot \mch^{\otimes k/2}$ (with $k$ even), then there is a projectively invariant state $\ket{\Psi'}$ which is maximally entangled across that partition, and so of the form $\ket{\Psi'}=d^{-k/4}\sum_{\boldsymbol{i},\boldsymbol{j}}\Om_{\boldsymbol{j},\boldsymbol{i}}\ket{\boldsymbol{i},\boldsymbol{j}}=(\id_{\mch^{\otimes k/2}}\ot\Om)\ket\Phi$ where $\Om$ is a projectively preserved unitary multilinear form on the Hilbert space $\mch^{\otimes k/2}$; that is, $U\tkt \Om (U^\mst)\tkt = \chi(U)\Om$.
Indeed, the extension to (even) $k>2$ proceeds by   just running the above arguments but with $\mch$ replaced by $\mch^{\otimes k/2}$: We use Lemma~\ref{lem:supp1} to find a (non-degenerate by the full rank of $\ket\Psi$) projectively preserved multilinear form, and then Lemma~\ref{lem:supp3} to upgrade it to a unitary projectively preserved multilinear form, the action of which on one half of a Bell pair gives us our required maximally entangled invariant state.\\

We now give a simple result (essentially, just the statement of the $G$-invariance of $\Om$ at the level of the Lie algebra of $G$) that will furnish us with a supply of groups to which our theorems will apply. 

\begin{prop}\label{lem:pauli}
Let $G$ be a group of unitaries of the form $U=\prod_j U_j=\prod_j e^{i\theta_j P_j}$, where $\theta_j\in\mbr$ and $P_j\in\mcs$ for a (group-defining) set of Pauli strings $\mcs$. Then $\Om$ is a $G$-invariant bilinear form if and only if, for each $P\in\mcs$, $P\Om + \Om P^\mst = 0$.
\end{prop}
\begin{proof}
First, note that for $\Omega$ to be a $G$-invariant bilinear form, it is certainly   necessary  for it to be preserved by each such $U_j$; in fact, this is also sufficient. Indeed, suppose $\Om$ is so preserved. Then we have
\begin{align}
    U\Om U^\mst &=\left(U_{N} U_{N-1} \cdots U_2 U_1 \right)\Omega  \left(  U_{N} U_{N-1} \cdots U_2 U_1 \right)^\mst \nonumber\\
     &=\left(U_{N} U_{N-1} \cdots U_2 U_1 \right)\Omega  \left( U_1^\mst U_2^\mst \cdots U_{N-1}^\mst U_{N}^\mst \right) \nonumber\\
    &= U_N  U_{N-1} \cdots U_{2}\left(  U_{1} \Omega U_{1}  ^\mst\right)U_{2}^\mst  \cdots U_{N-1}^\mst U_N^\mst \nonumber\\
    &= U_N  U_{N-1} \cdots U_{3} \left(  U_{2} \Omega U_{2} ^\mst \right)U_{3}^\mst \cdots U_{N-1} ^\mst U_N ^\mst \nonumber\\
    &\hspace{1.2mm} \vdots \nonumber\\
    &= U_N  \Omega U_N ^\mst\nonumber \\
    &=  \Omega \,.
\end{align}
So, let us focus (for some $P_j\in\mcs$) on $U_j\Om U_j ^\mst= e^{i\varepsilon P_j}\Om e^{i\varepsilon P_j^\mst}$. 
Now, if $P\Om+\Om P^\mst=0$, we immediately have $e^{i\varepsilon P_j}\Om e^{i\varepsilon P_j^\mst}=\Om e^{-i\varepsilon P_j^\mst} e^{i\varepsilon P_j^\mst}=\Om$. Conversely, suppose $e^{i\varepsilon P_j}\Om e^{i\varepsilon P_j^\mst}=\Om$ for all $\varepsilon\in\mbr$. Taking $\varepsilon\ll 1$ and
Taylor expanding the exponentials at the origin we have $\Om=e^{i\varepsilon P_j}\Om e^{i\varepsilon P_j^\mst} = \Omega + i\varepsilon (P_j\Om + \Om P_j^\mst) + \mco(\varepsilon^2)$, which immediately yields $P_j\Om + \Om P_j^\mst=0$. 
\end{proof}

In particular, recall that the matchgate group is generated by nearest-neighbor $XX$ and single qubit $Z$ gates. By the elementary anti-commutation properties of the Paulis we then have the following immediate corollary of Proposition~\ref{lem:pauli}:
\begin{restatable}{crl}{mgbf}
The matchgate group has the preserved bilinear forms $\Om_1 =XYXYX\ldots$ and $\Om_2=YXYXY\ldots$, and corresponding invariant states given by $(\id\ot\Om_j)\ket\Phi,\ j=1,2$. 
\end{restatable}
This establishes that our theorems apply to the matchgate group.

\section{Proof of main results}\label{sec:proofs}

In this section we prove Theorems~\ref{thm:main} and~\ref{thm:nlocal}.
We consider an $n$-qubit quantum system $\HC \coloneqq (\mbb{C}^2)\tn$ as a module for the action of some group $G\subseteq \mbu(\HC)$, with the goal of imposing restrictions on the depth and number of gates required for an ensemble over unitaries in $G$ to form designs over $G$. We do this by proving -- under various conditions -- that short-depth and low-gate-count ensembles can be distinguished from $k$-designs over $G$, and therefore cannot themselves be $k$-designs over $G$. Indeed, by explicitly choosing (not necessarily optimal or efficiently implementable) states and measurements, and computing a certain channel discrimination success probability, we are able to lower bound the distance between these restricted ensembles and $k$-designs over their respective groups.\\

\noindent
Let us recall:
\begin{algorithm}[Unitary channel discrimination]\label{algo}
Goal: Given two ensembles $\mc{E}$ and $\mc{F}$ of unitaries acting on $\HC$, and a state $\rho \in \LC(\HC^{\otimes k}\otimes \HC_{\rm anc})$, which is with probability $\lm$ evolved by a unitary  $U\sim \mc{E}$ and with probability $1-\lm$ evolved by a unitary  $U\sim \mc{F}$, determine which of these two possibilities occurred. Procedure:
    \begin{enumerate}
    \item Prepare a state $\rho \in \LC(\HC^{\otimes k}\otimes \HC_{\rm anc})$,
    \item Expose $\rho$ to black-box process that with probability $\lm$ applies $U\tk$ with $U\sim \mc{E}$ and with $1-\lm$ applies $U\tk$ with $U\sim \mc{F}$, 
    \item Implement a POVM $\Pi=\{\Pi_{\mc{E}},\Pi_{\mc{F}}\}$,
    \item Upon measuring $\Pi_\mc{X}$ for some $\mc{X}\in\{\mce,\mcf\}$, guess that $\rho$ was evolved by $U\sim\mc{X}$.
\end{enumerate}
\end{algorithm}

So, one guesses $\mc{E}$ with probability  $p(\Pi_\mce\hspace{0.7mm}|\hspace{0.7mm}U\sim\mcx)= \Tr[\rho_\mc{X} \Pi_{\mc{E}}]$, where $\rho_\mc{X} = (\phi^{(k)}_{\mc{X}} \otimes I_{\rm anc})(\rho)$. 
The probability of success in Algorithm~\ref{algo} is
\begin{align}
p_{\rm succ} (\rho,\Pi) &= p(U\sim\mce) p(\Pi_\mce\hspace{0.7mm}|\hspace{0.7mm}U\sim\mce)  + p(U\sim\mcf) p(\Pi_\mcf\hspace{0.7mm}|\hspace{0.7mm}U\sim\mcf) \nonumber\\
&=\lm \,p(\Pi_\mce\hspace{0.7mm}|\hspace{0.7mm}U\sim\mce)  + (1-\lm) \,p(\Pi_\mcf\hspace{0.7mm}|\hspace{0.7mm}U\sim\mcf)\,.
\end{align}
For any choice of $\rho$ and $\Pi$ producing a success probability better than random guessing, $p_{\rm succ} (\rho,\Pi)>1/2$, one obtains a non-trivial bound on the diamond distance between the channels given by (setting $\lm=1/2$)
\begin{equation}
    \|\phi^{(k)}_{\mc{E}}-\phi^{(k)}_{\mc{F}}\|_{\diamond} \geq 4(p_{\rm succ}(\rho,\Pi)-1/2)\,.
\end{equation}

\begin{figure*}[t!]
    \centering
    \includegraphics[width=1\linewidth]{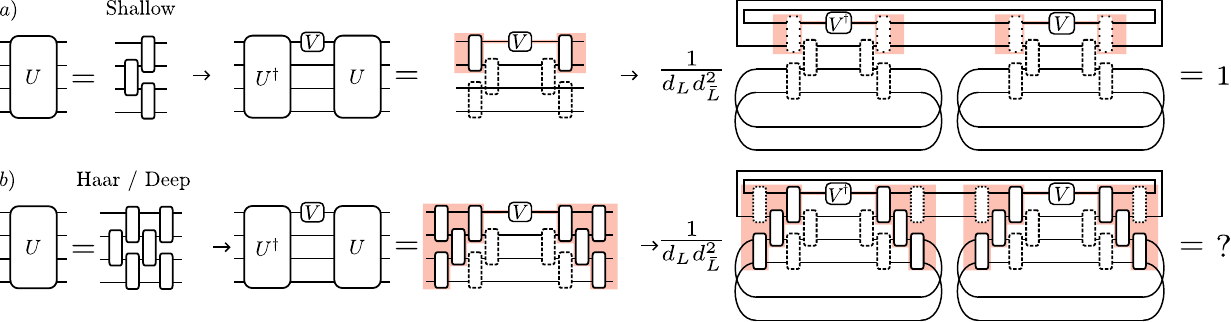}
    \caption{Given a subgroup $G$ of the unitary group that possesses a $G$-invariant state $\ket{\Psi}\in(\mch\tk)^G$ satisfying certain technical conditions (see Theorem~\ref{thm:main}),  our main result is an algorithm for distinguishing between $k$-designs over $G$ and \textit{any} ensemble of elements of $G$ consisting of sublinear-depth circuits. The key insight is that the existence of such a state implies the  
    ability to effectively time-reverse elements of $G$, allowing one  to distinguish between shallow and deep circuits by standard lightcone arguments. Specifically, we describe a single-shot experiment that with high probability distinguishes between deep and shallow ensembles, constructing a POVM element that is measured with probability one in the shallow case (top-right) and with vanishingly small probability in the $k$-design case (bottom-right). We can then use the connection of Eq.~\eqref{eq:succbound} between distinguishability and the diamond distance to rule out the existence of $k$-designs over $G$ that are composed of shallow elements of $G$. }
    \label{fig:POVM-ap}
\end{figure*}

\subsection{ Bounding circuit depth}

We begin by focusing on ensembles over quantum circuits with restricted depth. Let $\mc{E}_L$ be an arbitrary ensemble of depth-$L$ circuits belonging to $G\subseteq \mbu(\HC)$. By this we mean we consider circuits with $L$ time-steps, where each gate takes a single time-step, and gates acting on disjoint qubits can be performed simultaneously. We use $\mc{E}_G$ to denote the Haar ensemble over $G$,  and $\phi_L^{(k)}$ and $\phi_G^{(k)}$ to denote the corresponding average channels. We introduce a self-adjoint ``perturbation'' $V\in \mbu(\HC)$, $V=V\ad$, and consider the bipartition  $\mch  =\mch_L\otimes\mch_{\overline{L}}$, with $\mch_L$  being the Hilbert space associated to all the qubits over which, for any $U\sim\mce_L$,   $UVU^\dagger$ has support   (i.e., all the qubits over which $UVU^\dagger$ can act non-trivially). We refer the reader to Fig.~\ref{fig:POVM-ap}. 
We assume that there exists a projectively $G$-invariant state  of full Schmidt rank across $\mch^{\otimes k/2}\ot \mch^{\otimes k/2}$. From the discussion of Appendix~\ref{sec:forms} we know that we have a projectively $G$-invariant state of the form $\ket\Psi= (\id_{\mch^{\otimes k/2}}\ot\Om)\ket\Phi\in\mch\tk$, where $\Om$ is a unitary preserved multilinear form, and $\ket\Phi_\mch^{\ot k/2}$ is the Bell state between the two copies of $\mch^{\otimes k/2}$. We will denote by $d_L$ and $d_{\overline{L}}$ the dimensions of $\mch_L$ and $\mch_{\overline{L}}$, respectively. \\

Our strategy will be based on the observation that the state $\ket{\Psi_V}:=(V\otimes \id_{\mch^{\ot (k-1)}})\ket\Psi$ behaves detectably differently under the actions of $\mce_L$ and $\mce_G$; as we shall argue, this difference can be detected by the two-element POVM $\{\Pi_{\mce_L},\Pi_G\}$ given by
\begin{equation}
    \Pi_{\mce_L} = (\id_{\mch^{\ot k/2}}\ot\Om) \widetilde{\Pi_{\mce_L}}(\id_{\mch^{\ot k/2}}\ot\Om\ad);\qquad \Pi_G=\id_{\mch\tk}-\Pi_{\mce_L},
\end{equation}
where (writing basis vectors with respect to the refined decomposition $\mch\tk = (\mch_L\ot\mch_{\overline{L}})\tk$)
\begin{align}
    \widetilde{\Pi_{\mce_L}} &= d_{\overline{L}}^{-k/2}d_{{L}}^{-k/2+1}\sum_{\substack{i_2,i_3,\ldots i_{k/2}=1\\a_2,a_3,\ldots a_{k/2}=1}}^{d_L}\ \sum_{\substack{j_1,j_2,\ldots j_{k/2}=1\\b_1,b_2,\ldots b_{k/2}=1}}^{d_{\overline{L}}}\id_{\mch_L}\ot\ketbra{j_1 i_2j_2i_3j_3\ldots j_{k/2}}{b_1 a_2b_2a_3b_3\ldots b_{k/2}}\otimes\nonumber\\
    &\hspace{70mm}\id_{\mch_L}\ot   \ketbra{j_1 i_2j_2i_3j_3\ldots j_{k/2}}{b_1 a_2b_2a_3b_3\ldots b_{k/2}} \label{eq:piel}
\end{align}
In words: $\Pi_{\mce_L}$ corresponds to ``undoing'' the action of $\Om$, followed by ignoring the first copy of $\mch_L$ and contracting with Bell states everywhere else. In the notation of the above channel discrimination algorithm, what we would like to show is that, starting in the state $\ket{\Psi_V}$, both $p(\Pi_{\mce_L}\,|\,U\sim\mce_L)\approx 1$ and $p(\Pi_{G}\,|\,U\sim G)\approx 1$; we would then be able to distinguish with high confidence between unitaries sampled from the two ensembles, by applying (the $k$\textsuperscript{th} tensor power of) a given such unitary to $\ket{\Psi_V}$, performing the POVM, and seeing which outcome we get. To that end, we have a Lemma.
\begin{restatable}{lemma}{lems4}\label{lem:supp4}
The channel discrimination algorithm described above has $p(\Pi_{\mce_L}\,|\,U\sim\mce_L)= 1$ and
\begin{equation*}
    p(\Pi_{\mce_L}\,|\,U\sim\mce_G)=\expect_{U\sim \mce_G}  \overline{F}_{\mch_{\overline{L}}} (UVU\ad),
\end{equation*}
where $\overline{F}_{\mch_{\overline{L}}}(O)=d_{\overline{L}}^{-2}\sum_{P\,:\,{\rm supp}(P)\subset  \mch_{\overline{L}}} d^{-1}   \tr[O\ad P OP ] $ is the  infinite-temperature OTOC averaged over the Pauli probes $P$ with support on $\mch_{\overline{L}}$.
\end{restatable}

\begin{proof}
We begin by recalling that from the discussion of Appendix~\ref{sec:forms}, we know that $U^{\ot k/2}\Om=\chi(U)\Om (U^*)^{\ot k/2}$ for some character $\chi$ of $G$. In our present context, this implies
\begin{align}
    U\tk \ket{\Psi_V}&=(U\tkt (V\ot\id_{\mch}^{k/2-1}))\ot(U\tkt\Om)\ket{\Phi}\\
    &=\chi(U)(U\tkt (V\ot\id_{\mch}^{k/2-1}))\ot(\Om(U^*)\tkt)\ket{\Phi}\\
    &=\chi(U)((U\tkt (V\ot\id_{\mch}^{k/2-1})(U\ad)\tkt)\ot\Om)\ket{\Phi}\\
    &=\chi(U)((U VU\ad)\ot\id_{\mch}^{k/2-1}\ot\Om)\ket{\Phi}\label{eq:s12}
\end{align}
For example, for $k=4$ we would have in the graphical notation the calculation:

\begin{center}
\includegraphics[width=0.975\textwidth]{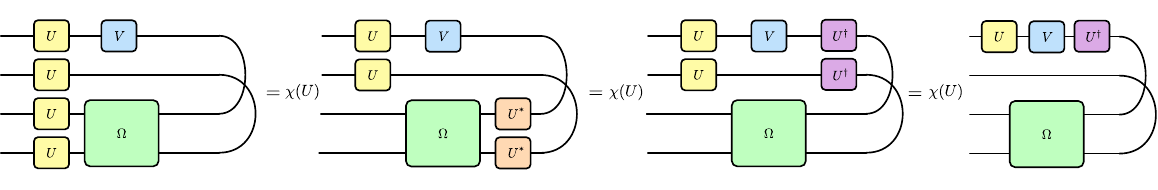}
\end{center}

\noindent
In agreement with Eq.~\eqref{eq:s12}. It follows that
\begin{align}
    p(\Pi_{\mce_L}\,|\,U\sim\mce_G) &= \expect_{U\sim\mce_G}\Tr[\Pi_{\mce_L}U\tk \ketbra{\Psi_V} (U\ad)\tk ]\\
    &=\expect_{U\sim\mce_G}\Tr[{\Pi_{\mce_L}}((UVU\ad)\ot\id_{\mch^{\ot (k/2-1)}}\ot\Omega) \ketbra{\Phi} ((UV\ad U\ad)\ot\id_{\mch^{\ot (k/2-1)}}\ot\Omega\ad) ]\\
    &=\expect_{U\sim\mce_G}\Tr[\widetilde{\Pi_{\mce_L}}((UVU\ad)\ot\id_{\mch^{\ot (k-1)}}) \ketbra{\Phi} ((UV\ad U\ad)\ot\id_{\mch^{\ot (k-1)}}) ]
\end{align}
Now, $\ketbra{\Phi}$ can be written 
\begin{align}
    \ketbra{\Phi} &= d_{\overline{L}}^{-k/2}d_{{L}}^{-k/2}\sum_{\substack{i_1,i_2,i_3,\ldots i_{k/2}=1\\a_1,a_2,a_3,\ldots a_{k/2}=1}}^{d_L}\ \sum_{\substack{j_1,j_2,\ldots j_{k/2}=1\\b_1,b_2,\ldots b_{k/2}=1}}^{d_{\overline{L}}} \ketbra{i_1j_1 i_2j_2i_3j_3\ldots j_{k/2}}{a_1b_1 a_2b_2a_3b_3\ldots b_{k/2}}\otimes\nonumber\\
    &\hspace{70mm}   \ketbra{i_1 j_1 i_2j_2i_3j_3\ldots j_{k/2}}{a_1b_1 a_2b_2a_3b_3\ldots b_{k/2}} \label{eq:phiel},
\end{align}
so that the contraction with $\widetilde{\Pi_{\mce_L}}$ can be read off from Eq.~\eqref{eq:piel}. Contracting over the $k-2$ copies of $\mch$ that involve neither the first copy nor the copy with which it is entangled simply yields a factor of $(d_Ld_{\overline{L}})^{k-2}=d^{k-2}$. It remains to perform the trace over the remaining two copies of $\mch$ (so that we have four kets: $\mch_L,\mch_{\overline{L}},\mch_L,\mch_{\overline{L}}$); we  obtain
\begin{align}
    p(\Pi_{\mce_L}\,|\,U\sim\mce_G) &= d^{k-2}(d_{\overline{L}}^{-k/2}d_{{L}}^{-k/2})(d_{\overline{L}}^{-k/2}d_{{L}}^{-k/2+1})\expect_{U\sim \mce_G}\sum_{i_1,a_1=1}^{d_L}\sum_{j_1,j_1',b_1,b_1'=1}^{d_{\overline{L}}  }{\rm Tr}\Big[ ((UVU\ad)\ot \id_{\mch})\ketbra{i_1j_1 i_1j_1}{a_1b_1 a_1b_1} \nonumber \\
    &\hspace{70mm}((UV\ad U\ad)\ot \id_{\mch})(\id_{\mch_L} \ot \ketbra{j_1'}{b_1'}\ot \id_{\mch_L} \ot \ketbra{j_1'}{b_1'})  \Big]\\
    &= \frac{d_L}{d^2} \expect_{U\sim \mce_G}\sum_{i_1=1}^{d_L}\sum_{j_1,b_1=1}^{d_{\overline{L}}  }{\rm Tr}\Big[ (UVU\ad)\ketbra{i_1j_1 }{i_1b_1} (UV\ad U\ad)(\id_{\mch_L} \ot \ketbra{b_1}{j_1} )\Big]\\
    &= \frac{d_L}{d^2} \expect_{U\sim \mce_G} {\rm Tr}_{\mch_L}\Big[ {\rm Tr}_{\mch_{\overline{L}}}[UVU\ad]  \,{\rm Tr}_{\mch_{\overline{L}}} [UV\ad U\ad]\Big]\label{eq:branch}
\end{align}
Now, let us expand $UVU\ad$ in the Pauli basis; that is, write $UVU\ad= d^{-1}\sum_P \Tr[P UVU\ad] P$.   Evidently the partial trace over $\mch_{\overline{L}}$ kills any Pauli which fails to act as the identity on $\mch_{\overline{L}}$ (and acts as the scalar $d_{\overline{L}}$ on the Paulis which do act trivially on $\mch_{\overline{L}}$) so that we have (using vertical lines to denote the restriction of an operator to a subspace of its domain)
\begin{align}
    p(\Pi_{\mce_L}\,|\,U\sim\mce_G) &= \frac{d_L}{d^4} \expect_{U\sim \mce_G} {\rm Tr}_{\mch_L}\Big[d_{\overline{L}}^2 \sum_{P\,:\,P\rvert_{\mch_{\overline{L}}}=\id_{\mch_{\overline{L}}}} \Tr[P UVU\ad] P\rvert_{\mch_{L}}  \sum_{Q\,:\,Q\rvert_{\mch_{\overline{L}}}=\id_{\mch_{\overline{L}}}} \Tr[Q UVU\ad]^* Q\rvert_{\mch_{L}} \Big]\\
    &=\frac{d_Ld_{\overline{L}}^2}{d^4} \expect_{U\sim \mce_G}  \Big[ \sum_{P\,:\,P\rvert_{\mch_{\overline{L}}}=\id_{\mch_{\overline{L}}}}\sum_{Q\,:\,Q\rvert_{\mch_{\overline{L}}}=\id_{\mch_{\overline{L}}}} \Tr[P UVU\ad]  \Tr[Q UVU\ad]^* \delta_{P,Q}d_L \Big]\\
    &= \frac{1}{d^2} \expect_{U\sim \mce_G}  \Big[ \sum_{P\,:\,P\rvert_{\mch_{\overline{L}}}=\id_{\mch_{\overline{L}}}} \Big| \Tr[P UVU\ad] \Big|^2 \Big]\label{eq:pproj}
\end{align}
Now, as the Paulis form an orthonormal basis of operator space with respect to the (normalized) Hilbert-Schmidt inner product $\sdbraket{A}{B}=d^{-1}\Tr[A\ad B]$, we see that Eq.~\eqref{eq:pproj} is exactly $\expect_{U\sim \mce_G}\sdbraket{UVU\ad}{\mcp_{\mch_L}(UVU\ad)} $, where $\mcp_{\mch_L}$ orthogonally projects onto the subspace of operators with support on $\mch_L$ (that is, those operators which act trivially on $\mch_{\overline{L}}$). 
Next, by virtue of the Paulis forming a unitary 1-design~\cite{mele2023introduction}, we also have that the average infinite-temperature OTOC
\begin{equation}
    \overline{F}_{\mch_{\overline{L}}}(O)=d_{\overline{L}}^{-2}\sum_{P\,:\,{\rm supp}(P)\subset  \mch_{\overline{L}}} d^{-1}   \tr[O\ad P OP ] 
\end{equation}
satisfies
\begin{align}
    \overline{F}_{\mch_{\overline{L}}}(O) &=  d^{-1}   \tr[O\ad \left( d_{\overline{L}}^{-2}\sum_{P\,:\,{\rm supp}(P)\subset  \mch_{\overline{L}}} P OP \right)]\\ 
     &=  d^{-1}   \tr[O\ad \mcp_{\mch_L} \left(  O \right)]\\ 
     &= \sdbraket{O}{\mcp_{\mch_L} \left(  O \right)}\,, 
\end{align}
whence the discussion of the previous paragraph immediately implies that
\begin{equation}
    p(\Pi_{\mce_L}\,|\,U\sim\mce_G) = \expect_{U\sim \mce_G} \overline{F}_{\mch_{\overline{L}}}(UVU\ad)\,.
\end{equation}

This establishes one of the two claims of the Lemma. For the second claim, we note that we can without obstruction repeat the above logic with the substitution $\mce_G\to\mce_L$; as by definition $UVU\ad$ acts trivially on $\mch_{\overline{L}}$ for $U\sim\mce_L$, however, we in this instance have
\begin{equation}
    p(\Pi_{\mce_L}\,|\,U\sim\mce_L) = \expect_{U\sim \mce_L}  \overline{F}_{\mch_{\overline{L}}} (UVU\ad)=\expect_{U\sim \mce_L}  1 = 1,
\end{equation}
establishing the second claim of the Lemma.

\end{proof}

\noindent
We can now prove Theorem~\ref{thm:main} (restated for convenience):
\begin{theorem}    \label{thm:k}
If there is a projectively $G$-invariant state in $\mch\tk$ of full Schmidt rank across $\mch\tkt\ot\mch\tkt$, then 
 \begin{equation}\label{eq:bound}
\|\phi_{\mce_L}^{(k)}-\phi_{\mce_G}^{(k)}\|_\diamond\geq 2-2\expect_{U\sim \mce_G}  \overline{F}_{\mch_{\overline{L}}} (UVU\ad)\,,
\end{equation}
where $\overline{F}_{\mch_{\overline{L}}}(O)=d_{\overline{L}}^{-2}\sum_{P\,:\,{\rm supp}(P)\subset  \mch_{\overline{L}}} d^{-1}   \tr[O\ad P OP ] $ is the  infinite-temperature OTOC averaged over the Pauli probes $P$ with support on $\mch_{\overline{L}}$.
\end{theorem}
\begin{proof}
Let us implement the above-described channel discrimination procedure. We take as the initial state the ``perturbed'' state $\ket{\Psi_V}$; which we imagined to be evolved by a unitary $U\tk$, where $U$ is with probability 1/2 sampled from $\mce_L$, and with probability 1/2 sampled from $\mce_G$. We measure with respect to the POVM $\{\Pi_{\mce_L},\Pi_{\mce_G}\}$. If we obtain the measurement outcome $\Pi_{\mce_L}$ (respectively $\Pi_{\mce_G}$) then we guess that $U$ was sampled from $\mce_L$ (respectively $\mce_G$). By Lemma~\ref{lem:supp4}, the probability that we guess correctly is given by
\begin{align}
p_{\rm succ} &=  p(\Pi_{\mce_L}\hspace{0.7mm}|\hspace{0.7mm}U\sim\mce_L)p(U\sim\mce_L) +p(\Pi_{\mce_G}\hspace{0.7mm}|\hspace{0.7mm}U\sim\mce_G)p(U\sim\mce_G) \\
&= \frac{  p(\Pi_{\mce_L}\hspace{0.7mm}|\hspace{0.7mm}U\sim\mce_L) + p(\Pi_{\mce_G}\hspace{0.7mm}|\hspace{0.7mm}U\sim\mce_G)}{2} \\
&= \frac{ p(\Pi_{\mce_L}\hspace{0.7mm}|\hspace{0.7mm}U\sim\mce_L) + ( 1- p(\Pi_{\mce_L}\hspace{0.7mm}|\hspace{0.7mm}U\sim\mce_G))}{2}\\
&= \frac{ 1}{2} \left(1 +  \Big( 1- \expect_{U\sim \mce_G}  \overline{F}_{\mch_{\overline{L}}} (UVU\ad) \Big)\right)
\end{align}
whence we have
\begin{align}
\|\phi_L^{(k)}-\phi_G^{(k)} \|_{\diamond}  &\geq 4(p_{\rm succ}-1/2)\\
&= 2- 2\expect_{U\sim \mce_G}  \overline{F}_{\mch_{\overline{L}}} (UVU\ad)\,,
\end{align}
which is the desired expression.    
\end{proof}

\noindent
When doing examples we will also make use of the following alternative formulation:
\setcounter{crl}{10}
\begin{crl}
If there is a projectively $G$-invariant state in $\mch\tk$ of full Schmidt rank across $\mch\tkt\ot\mch\tkt$, then 
 \begin{equation}\label{eq:swapbound}
\|\phi_{\mce_L}^{(k)}-\phi_{\mce_G}^{(k)}\|_\diamond\geq 2-\frac{2 d_L}{d^2}\Tr[\phi_G^{(2)} \left(V\ot V^\dagger\right)\mss_L]\,,
\end{equation}
where   $\mss_L$ is the swap operator on the two copies of $\mch_L$ in $\mch\ts$. 
\end{crl}
\setcounter{crl}{1}
\begin{proof}
    Based on our above results, it clearly suffices to prove 
    \begin{equation}
        p(\Pi_{\mce_L}\,|\,U\sim\mce_G) = \frac{ d_L}{d^2}\Tr[\phi_G^{(2)} \left(V\ot V^\dagger\right)\mss_L].
    \end{equation}
    Indeed, picking up from Eq.~\eqref{eq:branch} we have
    \begin{align}
        p(\Pi_{\mce_L}\,|\,U\sim\mce_G) &= \frac{d_L}{d^2} \expect_{U\sim \mce_G} {\rm Tr}_{\mch_L}\Big[ {\rm Tr}_{\mch_{\overline{L}}}[UVU\ad]  \,{\rm Tr}_{\mch_{\overline{L}}} [UV\ad U\ad]\Big]\\
        &=\frac{d_{{L}}}{d^2}\expect_{U\sim \mce_G}\Tr[U\ts\left(V\ot V^\dagger\right)(U\ad)\ts\mss_L]\\
    &=\frac{d_{{L}}}{d^2}\Tr[\phi_{\mce_G}^{(2)}\left(V\ot V^\dagger\right)\mss_L]\label{eq:s36}
\end{align}
where we recall that $\mss_L$ is a swap operation on $\mch_L \td \subset \HC\td$ (i.e., for any $\ket{\Psi_{L}},\ \ket{\varphi_{L}}\in\mch_L$ and  $\ket{\Psi_{\overline{L}}},\ \ket{\varphi_{\overline{L}}}\in\mch_{\overline{L}}$,  $\mss_L\ket{\Psi_L}\ket{\Psi_{\overline{L}}}\ket{\varphi_L}\ket{\varphi_{\overline{L}}}=\ket{\varphi_L}\ket{\Psi_{\overline{L}}}\ket{\Psi_L}\ket{\varphi_{\overline{L}}}$, extended linearly to a map on $\mch\ts$). Graphically, this short calculation can be understood within the greater context of Lemma~\ref{lem:supp4}  as the manipulation
\begin{center}
    \includegraphics[width=\textwidth]{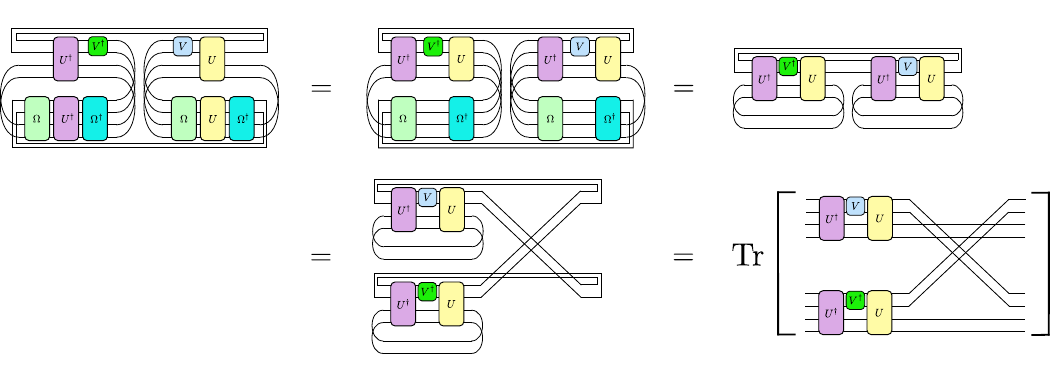}
\end{center}
To deduce the statement of the Corollary from Eq.~\eqref{eq:s36} is to repeat the argument of the proof of Theorem~\ref{thm:k}.
\end{proof}

\subsection{Bounding gate count}

\begin{figure}[t!]
    \centering
    \includegraphics[width=0.4\linewidth]{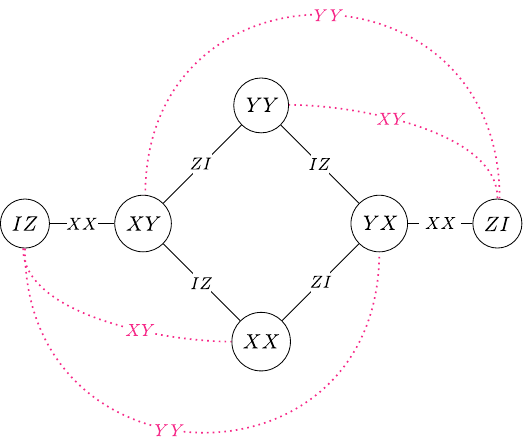}
    \caption{A connected component of the \textit{commutator graph}~\cite{diaz2023showcasing,west2025graph} corresponding to the generating set $\{ZI,IZ,XX\}$ (solid lines) optionally augmented with the generators $\{XY,YY\}$ (dotted lines). Each vertex of the graph is a Pauli string; the edges (present between vertices linked by the adjoint action of a generator) give a generating-set-dependent notion of  locality in \textit{operator space}. We exploit this in Theorem~\ref{thm:nlocal} to rule out short-depth designs over a class of ensembles of unitaries not necessarily covered by Theorem~\ref{thm:main}. }
    \label{fig:cgraph}
\end{figure}

\noindent
Before coming to the proof of Theorem~\ref{thm:nlocal}, we establish some notation. Consider any ensemble $\mce_N^\mcs$ over elements of the form $U=\prod_{\ell=1}^N \exp(i\th_\ell H_\ell)$, where the generators are Paulis $\{H_\ell\}_\ell\subset\mcs\subset\mfg$. Here,  $\mfg$ denotes the Lie algebra associated to the Lie group $G$ generated by $\SC$, and is assumed to possess a preserved bilinear form $\Omega$ (i.e. for all $U\in G,\ U\Om U^\mst = \Om$) and an invariant state $\ket{\Psi}\in\mch\ts$. As  discussed in Appendix~\ref{sec:forms}, we can write such a state as $\ket\Psi = \id\ot\Omega\ket\Phi$, where $\ket\Phi$ is a Bell state across the two copies of $\mch$. Fix a Pauli $P$.  With respect to the choice $\mcs$ of generating set, we have a commutator graph $\mcg_{\mcs}$ (recalling Definition~\ref{def:cg} and seeing Fig.~\ref{fig:cgraph}). Let $\mcn_{N,\mcs}(P)$ be the $N$-neighborhood of $P$ in $\mcg_{ \mcs}$, and $\mcn_{\mcs}(P)$ the connected component of $\mcg_{ \mcs}$ containing $P$.  Define projectors $\Pi_N, \Pi_{\overline{N}},\Pi_R$ onto (respectively) 
\begin{equation}
    \mch_N={\rm span}\{(T\ot\id)\ket{\Phi}\ :\ T\in \mcn_{N,\mcs}(P)\}\,,\quad 
    \mch_{\overline{N}}={\rm span}\{(T\ot\id)\ket{\Phi}\ :\ T\in \mcn_{\mcs}(P) {\rm\ but\ } T\not\in \mcn_{N,\mcs}(P)\}\,,
\end{equation}
and $\mch_R$ the orthogonal complement of $\mch_N\oplus\mch_{\overline{N}}$. Importantly, note that the $(T\ot\id)\ket{\Phi}$ are orthogonal for different choices of $T$. Let us now prove:
\begin{lemma}
    If $U\in\mce_N^{\mcs}$, then $(\id\ot \Om^\dagger)U\ts (P\ot\id)\ket{\Psi}\in\mch_N$. 
\end{lemma}
\begin{proof}
By our usual calculation, the invariance of $\ket{\Psi}=(\id\ot \Om)\ket{\Phi}$ allows us to write
\begin{equation}
(\id\ot \Om^\dagger)U\ts (P\ot\id)\ket{\Psi}=(\id\ot \Om^\dagger) (UPU^\dagger\ot\id)(\id\ot \Om)\ket{\Phi}=(UPU^\dagger\ot\id)\ket{\Phi}\,.
\end{equation}
Now, 
\begin{equation}
UPU^\dagger=U_1\cdots U_N P U_N^\dagger\cdots U_1\ad =  \exp(i\th_1H_1)\cdots  \exp(i\th_NH_N) P  \exp(-i\th_NH_N)\cdots \exp(-i\th_1H_1);
\end{equation}
the key point is that 
\begin{equation}
    \begin{cases}
        \exp(i\th_NH_N) P  \exp(-i\th_NH_N) = P\quad &{\rm if\ }[P,H_N]=0\\
        \exp(i\th_NH_N) P  \exp(-i\th_NH_N) = \cos(2\th_N)P+i\sin(2\th_N)H_NP\quad &{\rm if\ }\{P,H_N\}=0\\
    \end{cases}
\end{equation}
so that, either way, the action of $\exp(\pm i\th_NH_N)$ evolves $P$ into a linear combination of Pauli strings that are  a distance of at most one from $P$ on the commutator graph.  At the next step, $\exp(i\th_{N-1}H_{N-1})$ evolves $P$ and (maybe) $H_NP$  into linear combinations of themselves and Paulis adjacent to them on the graph; iterating, we see that the final state lies in $\mch_N$. 
\end{proof}

\noindent
We can now prove Theorem~\ref{thm:nlocal} (restated for convenience)
\setcounter{theorem}{1}
\begin{theorem} 
Let $\mce_N^\mcs$ be an ensemble of circuits  of the form $U=\prod_{\ell=1}^N\exp(i\vartheta_\ell H_\ell)$, where  $\{H_\ell\}_\ell\subset\mcs$ are Pauli strings in the Lie algebra of a Pauli-compatible, bilinear-form-preserving group $G$. Then, 
 \begin{equation}\label{eq:sthm2}
\|\phi_{\mce_N^\mcs}^{(2)}-\phi_{\mu_{G}}^{(2)}\|_\diamond\geq 2-2\frac{\left\lvert \mcn_{N,\mcs}(P)\right\rvert}{\left\lvert \mcn_{\mcs}(P)\right\rvert},
\end{equation}
where $\mcn_{N,\mcs}(P)$ denotes the set of vertices within a distance of at most $N$ of $P$ in the commutator graph corresponding to the generating set $\mcs$, and $\mcn_{\mcs}(P)$ the connected component of the graph containing $P$.
\end{theorem}
\begin{proof}
The proof is conceptually similar to that of Theorem~\ref{thm:main}. 
Fix a Pauli $P$, and consider the  commutator graph $\mcg_{\mcs}$. Let $\mcn_{N,\mcs}(P)$ be the $N$-neighborhood of $P$ in $\mcg_{ \mcs}$, and $\mcn_{\mcs}(P)$ the connected component of $\mcg_{ \mcs}$ containing $P$.  Define  projectors $\Pi_N, \Pi_{\overline{N}}$ onto (respectively) 
\begin{equation}
    \mch_N={\rm span}\{(T\ot\id)\ket{\Phi}\ :\ T\in \mcn_{N,\mcs}(P)\}
\end{equation}
and
\begin{equation}
    \mch_{\overline{N}}={\rm span}\{(T\ot\id)\ket{\Phi}\ :\  T\not\in \mcn_{N,\mcs}(P)\}.
\end{equation}
Importantly, note that the $(T\ot\id)\ket{\Phi}$ are orthogonal for different choices of $T$. We now have the
following modification of the experiments of Theorem~\ref{thm:main}: prepare $(P\ot\id)\ket{\Psi}$, evolve with either $\mce_N^{\mcs}$ or $\mu_{G}$, undo $\Om$, measure the POVM $\{\Pi_N, \Pi_{\overline{N}}\}$. 
Our strategy will be to guess $\mce_N^{\mcs}$ if we measure $\Pi_N$, and $\mu_{G}$ if we measure $\Pi_{\overline{N}}$. By the previous discussion, we are wrong if and only if the ensemble is $\mu_{G}$ and we measure $\Pi_N$. Therefore, with $d=2^n$,
\begin{align}
\|\phi_{\mce_N^\mcs}^{(2)}-\phi_{\mu_{G}}^{(2)}\|_\diamond&\geq 2 - 2 \expect_{U\sim\mu_{G}}\Tr[(UPU^\dagger\ot\id)\ketbra{\Phi}(UPU^\dagger\ot\id)\Pi_N]\nonumber\\
&= 2 - 2 \expect_{U\sim\mu_{G}}\Tr[(UPU^\dagger\ot\id)\ketbra{\Phi}(UPU^\dagger\ot\id)\sum_{T\in \mcn_{N,\mcs}(P)}(T\ot\id)\ketbra{\Phi}(T\ot\id)]\nonumber\\
&= 2 - \frac{2}{d^2} \sum_{T\in \mcn_{N,\mcs}(P)}\expect_{U\sim\mu_G}\Tr[(UPU^\dagger T)\ts]\label{eq:b25}\\
&= 2 - \frac{2}{d^2} \sum_{T\in \mcn_{N,\mcs}(P)} \sum_{j,\kappa} \Tr[Q_{j,\kappa}P\ts]\Tr[Q_{j,\kappa}T\ts]\label{eq:b26}\\
&= 2 - \frac{2}{d^2} \sum_{T\in \mcn_{N,\mcs}(P)} \sum_{j,\kappa} \left(\frac{1}{d\sqrt{|C_\kappa|}}\right)^2 \sum_{S,R\in C_\kappa} \Tr[(S\ot L_j S)P\ts]\Tr[(R\ot L_j R)T\ts]\label{eq:b27}\\
&= 2 - \frac{2}{d^2} \sum_{T\in \mcn_{N,\mcs}(P)} \sum_{j,\kappa} \frac{\delta_{L_j,\id}\delta_{C_\kappa, \mcn_{\mcs}(P)}}{d^2|C_\kappa|} d^4\label{eq:b28}\\
&= 2 \left(1 - \frac{\left\lvert \mcn_{N,\mcs}(P)\right\rvert}{\left\lvert \mcn_{\mcs}(P)\right\rvert}\right) \label{eq:nlexact} 
\end{align}
which is what we wanted to show. In going from Eq.~\eqref{eq:b25} to Eq.~\eqref{eq:b26} we have used the well-known fact that the expectation value in Eq.~\eqref{eq:b25} projects onto the quadratic symmetries of $G$; that is, the subspace of operators $O\in\mcl(\mch\ts)$ satisfying
$[U\ts, O]  = 0$
for all $U\in G$. Importantly, a (2-norm orthonormalized) basis for this space is known for any  Pauli-compatible group (with, say, a Pauli basis $\{L_j\}_{j=1}^J$ of its linear symmetries); it is given by the span of the operators
\begin{equation}
    Q_{j,\kappa} = \frac{1}{d \sqrt{\lvert C_\kappa\rvert}}\sum_{S\in C_\kappa} S\ot (L_jS)\,,
\end{equation}
where $C_\kappa$ denotes a connected component of the \textit{commutator graph} of $\mcs$~\cite{diaz2023showcasing,west2025graph}.
For some Pauli string $P$, 
we clearly have $\Tr[Q_{j,\kappa}^\dagger P\ts]\neq 0$ if and only if  $C_\kappa$ is the connected component of the commutator graph containing $P$, and $L_j=\id$; this observation underlies our going from Eq.~\eqref{eq:b27} to Eq.~\eqref{eq:b28}.
\end{proof}

\noindent
We remark that from Eq.~\eqref{eq:nlexact}  one  immediately obtains the simple  (but in general very loose) bound
\begin{equation}\label{eq:nlbound}
\|\phi_{\mce_N^\mcs}^{(2)}-\phi_{\mu_{G}}^{(2)}\|_\diamond \geq 2\left(1 - \frac{\left\lvert \mcs\right\rvert^N+1}{\left\lvert \mcn_{\mcs}(P)\right\rvert} \right).
\end{equation}

\section{Details of examples}\label{sec:exampledetails}
In this section we give the details of the examples of our formalism applied to specific groups considered in the main text.

\subsection{Pauli-compatible groups}\label{sec:pauli}
In this section we derive Eq.~\eqref{eq:boundpauli} by evaluating Eq.~\eqref{eq:swapbound} in the case of a Pauli-compatible group generated by a set $\mcs$ of Pauli strings. 
We can evaluate the expectation value in Eq.~\eqref{eq:swapbound} using the basis of quadratic symmetries provided in the proof of Theorem~\ref{thm:nlocal} (in Appendix~\ref{sec:proofs}).
In particular, for some Pauli string $P$, 
with $C(P)$ denoting its connected component of the commutator graph $\mcg_\mcs$ containing $P$, we from that discussion see
\begin{align}
\|\phi_{\mce_L}^{(2)}-\phi_{\mce_G}^{(2)}\|_\diamond &\geq 2-\frac{2}{dd_{\overline{L}}}\Tr[\expect_{U\sim\mce_G}(UPU^\dagger)\ts\mss_L]\nonumber\\
&=2-\frac{2}{dd_{\overline{L}}}\Tr[\sum_{j,\kappa} \Tr[Q_{j,\kappa}P\ts] Q_{j,\kappa}\mss_L]\nonumber\\
&=2-\frac{2}{d_{\overline{L}}\sqrt{|C(P)|}}\Tr[   Q_{\id,C(P)}\mss_L]\nonumber\\
&=2-\frac{2}{dd_{\overline{L}}|C(P)|}\sum_{T\in C_{P}} \Tr[T\ts \mss_L]\nonumber\\
&=2-\frac{2}{dd_{\overline{L}}|C(P)|}\sum_{T\in C_{P}} d_L d_{\overline{L}}^2\delta_{T_{\overline{L}}=\id_{\overline{L}}}\nonumber\\
&=2(1-r_G(P,L))\,,\label{eq:pbound}
\end{align}
where $r_G(P,L)$ is the fraction of Pauli strings in  $C(P)$  that act as the identity on $\mch_{\overline{L}}$. This proves Corollary~\ref{crl:pauli}. For example, let us suppose that the Lie algebra  has an ideal of dimension $4^{n-k}$ for some constant $k\ll n$, an assumption that is true in many cases of interest~\cite{wiersema2023classification}. If this ideal has a basis of Pauli strings, it will appear as a single connected component $C$ of the commutator graph of size $4^{n-k}$~\cite{west2025graph}. As there are at most $4^L$ strings that act as the identity on $\overline{L}$ within $C$, taking $L=n/2$ we obtain (for any $P\in C$), 
$r_G(P,n/2)\leq 4^{n/2}/4^{n-k}=\mco(2^{-n})$. 

\subsection{Matchgate group}\label{sec:mgate}
Having established in the Appendix~\ref{sec:forms} the  existence of a matchgate-invariant bilinear form, we are in a  position to apply Theorem~\ref{thm:main}, and therefore wish to evaluate the right hand side of the bound Eq.~\eqref{eq:bound} (in fact, we will use Eq.~\eqref{eq:pbound}).
Let us take the local perturbation to be a Pauli $X$ gate on qubit $n/2$ (taking, say, $n$ to be even; the odd case is similar), and  further take  $\mch_L$ to be the first $n-1$ qubits, and $\mch_{\overline{L}}$ the last qubit. As the matchgate group possesses a Pauli Lie algebra (and indeed a basis $\{\id,Z\tn\}$ of Pauli strings of the span of its linear symmetries) we can evaluate  the expectation value via Weingarten calculus~\cite{mele2023introduction} as in the previous section. In order to carry out the counting required to evaluate $r_{{\rm Spin}(2n)}(X_{n/2},n/2-1)$, however, we first need to recall some details of the matchgate group. To that end, recall that the matchgate Lie algebra is generated by the products of two \textit{Majorana fermions}, the set of which we can take to be enumerated as
\begin{alignat}{3}\label{eq:majos}
c_1&=XIII\cdots I,\qquad&c_{2}&=YIII\cdots I\nonumber\\
c_3&=ZXII\cdots I,&c_{4}&=ZYII\cdots I\\
&\:\:\vdots&&\:\:\vdots\nonumber\\
c_{2n-1}&=ZZ\cdots ZX,&c_{2n}&=ZZ\cdots ZY\;.  \nonumber
\end{alignat}
One can show~\cite{diaz2023showcasing}   that the connected components  $C_\kappa$ of the matchgate commutator graph are exactly given by the set of the possible products of $\kappa$ distinct Majoranas, for $0\leq \kappa\leq 2n$, and that furthermore 
$X_{n/2}$ may be written as the product of $n-1$ Majoranas\footnote{Note this has nothing to do with  $\mch_L$ being the first $n-1$ qubits}~\cite{diaz2023showcasing}. From our expression Eq.~\eqref{eq:majos} for the Majoranas, we see that a product of $n-1$ of them is the identity on $\mch_{\overline{L}}$ (i.e., the last qubit) if and only if the Majoranas which appear in the product have indices less than or equal to $2n-2$. Recalling Stirling's approximation, $n!\sim \sqrt{2\pi n}(n/e)^n$, we therefore see
\begin{align}
\|\phi_{\mce_L}^{(2)}-\phi_{\mu_{{\rm Spin}(2n)}}^{(2)}\|_\diamond &\geq 2-2r_{{\rm Spin}(2n)}(X_{n/2},n/2-1)\nonumber\\
&=2-2{2n-2 \choose n-1}{2n \choose n-1}^{-1}\nonumber \\
&=2-2\frac{n(n+1)}{2n(2n-1)}\nonumber\\
&\sim \frac{3}{2}\,.
\end{align}
As the support of $X_{n/2}$ is contained within $\mch_L$ for 2-qubit-generated matchgates up to (at least) a depth of $n/2-1$, we conclude
\setcounter{crl}{0}
\begin{crl}
Any ensemble of matchgate circuits requires depth $L\geq n/2$ to form an approximate matchgate $2$-design.
\end{crl}
Next we consider the application of Theorem~\ref{thm:nlocal}. As above, the analysis will be facilitated by the fact that the commutator graph is very well-behaved. This time we will focus our efforts on the largest component, $\kappa=n$. 
We begin by noting that, if the generating set is the usual $\{Z_i,X_iX_{i+1}\}$ then the diameter of $C_n$ is ${\rm diam}(C_{G;\{Z_i,X_iX_{i+1}\}}(P\in C_n))=n^2$~\cite{west2025graph}; at the other extreme we have ${\rm diam}(C_{G;\{{\rm all\ paulis\ in\ }\mfg\}}(P\in C_n))=n$. 
To see this second assertion, note that (with $a_i<a_j$ and $b_i<b_j$ for all $i<j$) to walk across the graph from the vertex $c_{a_1}c_{a_2}\cdots c_{a_n}$ to the vertex $c_{b_1}c_{b_2}\cdots c_{b_n}$ takes $n-\lvert\{c_{a_i}\}_i\cap\{c_{b_i}\}_i\rvert$   steps; this can be up to $n$ (note that you cannot change two indices at once, for any generator that does it commutes with the Pauli). These results suggest that, for these two choices of generating sets, Theorem~\ref{thm:nlocal} may be able to provide non-trivial bounds for, respectively, $N<n^2, n$. Before undertaking a careful analysis, we note that for $N\sim\log n$,
 Eq.~\eqref{eq:nlbound}  leads to
\begin{equation}
    \|\phi_{\mce_{N\sim\log n}^\mcs}^{(2)}-\phi_{\mu_{G}}^{(2)}\|_\diamond  \gtrsim 2- \frac{2n^{a\log n}}{4^n/\sqrt{\pi n}}\approx 2
\end{equation}
where $a=1,2$ for $S=\{XX,Z\},\ \{{\rm all\ paulis\ in\ }\mfg\}$ respectively. So, there are no approximate matchgate 2-designs over an ensemble of circuits that consist of products of $\sim\log n$ unitaries of the form $\exp(i\th H)$, for any Pauli $H$ in the matchgate DLA. In fact, we will now see that one can improve the $n$-dependence of this statement exponentially. 

\subsubsection{Full generating set}
In this instance the big component $C_n$ is isomorphic to the \textit{Johnson graph} $J(2n,n)$: the vertices are labelled by the $n$–element subsets of a $2n$–element set, with two vertices  adjacent iff $n-1$ of their elements are the same. 
The neighborhood of a given $P\in C_n$ (all vertices in a given component look the same) is therefore of size
\begin{equation}
\left\lvert \mcn_{N,\{{\rm all\ paulis\ in\ }\mathfrak{spin}(2n)\}}(P)\right\rvert = \sum_{k=0}^N{n \choose k}^2.
\end{equation}
This is because at each step, we pick one of the Majoranas in the decomposition of the Pauli corresponding to the vertex and replace it with another Majorana. There are (in $C_n$) $n$ choices of which guy to replace, and $2n-n=n$ choices of replacement. 
So, we care to bound (take $N=n/c$, say, for any $c>2$)
\begin{align}
\frac{\left\lvert \mcn_{n/c,\{{\rm all\ paulis\ in\ }\mathfrak{spin}(2n)\}}(P)\right\rvert}{\left\lvert C_{\{{\rm all\ paulis\ in\ }\mathfrak{spin}(2n)\}}(P)\right\rvert}&= {2n \choose n}^{-1}\sum_{k=0}^{n/c}{n \choose k}^2\nonumber\\
&\leq (n/c+1){2n \choose n}^{-1}{n \choose n/c}^2\,.
\end{align}
Applying Stirling's approximation, $n!\sim\sqrt{2\pi n}(n/e)^{n}$, then yields 
\begin{equation}
\frac{\left\lvert \mcn_{n/c,\{{\rm all\ paulis\ in\ }\mathfrak{spin}(2n)\}}(P)\right\rvert}{\left\lvert C_{\{{\rm all\ paulis\ in\ }\mathfrak{spin}(2n)\}}(P)\right\rvert}\lesssim \frac{c(n+c)}{2(c-1)\sqrt{\pi n}} \left( \frac{c}{2(c-1)^{1-1/c}}\right)^{2n}.
\end{equation}
An elementary calculation shows that $f(c)=c(c-1)^{1/c-1}/2$ is less than one for all $c>2$; for any fixed $c>2$ by Eq.~\eqref{eq:sthm2} we therefore have
 \begin{equation}
\|\phi_{\mce_{N=n/c}}^{(2)}-\phi_{\mu_{G}}^{(2)}\|_\diamond\gtrsim 2-\mco\left(\sqrt{n}b^{-n}\right)\,,
\end{equation}
for some $b>1$. This proves Corollary~\ref{crl:mgarb}.

\subsubsection{Standard generating set}
Next, let us consider the case of taking the generating set to contain only $\{Z_i,X_iX_{i+1}\}$, i.e., standard matchgate circuits. Focusing again on the component $C_n$, we therefore have a subgraph of $J(2n,n)$, with the same set of vertices but many of the edges now missing. Amongst other consequences, this increases the diameter of the component from $n$ to $n^2$~\cite{west2025graph}; for example, the vertices corresponding to the Pauli strings $P=c_1c_2\cdots c_n$ and $P'=c_{n+1}\cdots c_{2n}$ are separated by a distance of $n^2$. More generally, we see that by the symmetry enforced by the graph automorphism $Z\tn:C_n\to C_n$ that $\mcn_{n^2/2-1,\{Z_i,X_iX_{i+1}\}}/\abs{C_{\{Z_i,X_iX_{i+1}\}}}(P)< 1/2$, implying that for $N=n^2/2-1$ the experiment of Theorem~\ref{thm:nlocal} succeeds with probability $>0.75$, and therefore that in this setting $\|\phi_{\mce_{N=n^2/2-1}}^{(2)}-\phi_{\mu_{{\rm Spin}(2n)}}^{(2)}\|_\diamond\geq 1$, whence: 
\setcounter{crl}{6}
\begin{crl}\label{crl:mggsupp}
Any ensemble over 2-local matchgate circuits consisting of less than  $n^2/2$ gates can be efficiently distinguished from a matchgate 2-design.
\end{crl}

Next we turn to the proof of our result concerning the spectral gap of matchgate 2-expanders. To that end, we begin by recalling that we call an ensemble $\mce$ of unitaries in $G$ a $\lm$-approximate $t$-expander over $G$ if 
\begin{equation}\label{eq:texpsupp}
    \| \phi^{(t)}_\mce - \phi^{(t)}_{\mu_G}\|_{\infty} \le \lm
\end{equation}
for  $\lm<1$, and define the associated \textit{spectral gap} to be $\Delta:=1-\lm$. We now prove:
\begin{crl}\label{crl:mgexpsupp}
Any matchgate 2-expander whose one-step channel uses $\mco (1)$ elementary matchgates has $\Delta\in\mco(1/n)$.
\end{crl}
\begin{proof}
Let $\mce$ be a $\lm$-approximate $t$-expander over the $n$-qubit matchgate group, with each unitary of $\mce$ consisting of $g(n)$ elementary matchgates, and denote by $\mce^m$ the ensemble generated by composing $\mce $ with itself $m$ times. 
Having recalled both that $\mce^m$ is a $\lm^m$-approximate $t$-expander over $G$~\cite{mele2023introduction}, and the readily verified relation $\|\Phi\|_\diamond \le d^k\|\Phi\|_\infty$ for any superoperator $\Phi \in {\rm End}\,\mch\tk$ on a $d$-dimensional Hilbert space $\mch$,  the proof consists simply of some algebraic manipulations. Indeed, for $m < n^2/(2g(n))$ the ensemble $\mce^m$ consists of unitaries formed from the product of less than $n^2/2$ elementary matchgates, so that from the discussion preceding Corollary~\ref{crl:mggsupp} we have $\| \phi^{(2)}_{\mce^m} - \phi^{(2)}_{\mu_G}\|_{\diamond} \ge 1 $, whence 
\begin{equation}
    1\le d^2 \|\phi^{(2)}_{\mce^m} - \phi^{(2)}_{\mu_G}\|_{\infty}\le d^2 \lm^m = 4^n\lm^m\,.
\end{equation}
Using $\Delta=1-\lm$ then yields
\begin{equation}
 -\log(1-\Delta) \le \frac{n}{m}\log 4  \,. 
\end{equation}
Taking $m\sim n^2/(2g(n))$ and $g(n)\in\mco(1)$, the Taylor expansion $-\log(1-\Delta)\approx \Delta$ then gives $\Delta\in\mco(1/n)$, as required. 
\end{proof}

\subsection{The orthogonal and symplectic groups}\label{sec:ortho}
\noindent
In this section we apply Theorem~\ref{thm:main} to the orthogonal and symplectic groups,
\begin{equation}\label{eq:os}
    \mbb{O}(d) = \{U\in\mbu(d)\ :\ U\Om_{\mbo} U^\mst = \Om_{\mbo}\} \quad \text{ and }\quad\mbsp(d/2)=\{U\in\mbu(d)\ :\ U \Om_{\mbsp} U^\mst = \Om_{\mbsp}\},
\end{equation}
where we take $\Om_\mbb{O}=\id$ and $\Om_{\mbb{SP}} = iY_1$~\cite{garcia2024architectures,west2024random}. Note that from Eq.~\eqref{eq:os} both the orthogonal and symplectic groups manifestly preserve a bilinear form, so that by Appendix~\ref{sec:forms} they possess a suitable invariant state in $\mch\ts$, and Theorem~\ref{thm:main}  applies. 
In both cases, we take the groups to be generated by 2-qubit gates, and the perturbation to be $V=Z_1$, the Pauli-$Z$ operator on the first qubit. We now appeal to the Weingarten calculus on the orthogonal and symplectic groups~\cite{hashagen2018real,west2024real,collins2006integration,west2025graph,collins2009some} to conclude that, in either case,
\begin{equation}\label{eq:ortholin}
\expect_{U\sim\mce_G}\left[\left(UZ_1U\ad  \right)^{\otimes 2}\right]=\a_G\id\td+\beta_G\hspace{0.3mm}\mss+\gamma_G (\id\otimes\Omega_G)\ketbra{\Phi}(\id\otimes\Omega_G)\ad\,.
\end{equation}
for some to-be-determined coefficients $\a_G,\b_G$ and $\g_G$.  Multiplying both sides of Eq.~\eqref{eq:ortholin} by $\id\td,\ \mss$ and $ (\id\otimes\Omega_G)\ketbra{\Phi}(\id\otimes\Omega_G)\ad$ and taking traces yields a set of three simultaneous linear equations, whose solutions are
\begin{equation}
    \a_\mbo= \frac{-2}{(d+2)(d-1)}\,,\quad \b_\mbo=\frac{d}{(d+2)(d-1)} ,\quad \text{and}\quad \g_\mbo=\frac{d^2}{(d+2)(d-1)} \,,
\end{equation}
and
\begin{equation}
    \a_\mbsp= 0\,,\quad \b_\mbsp=\frac{1}{d+1} ,\quad \text{and}\quad \g_\mbsp=-\frac{d}{d+1} \,,
\end{equation}
Inserting this into Eq.~\eqref{eq:swapbound} we obtain in both cases that
\begin{align*}
\|\phi_{L}^{(2)}-\phi_{G}^{(2)}\|_\diamond&\geq 2-\frac{2d_L}{d^2}\Tr[\phi_{G}^{(2)}\left(Z_1\ts\right)\mss_L]\approx 2-2\left(\frac{d_L}{d}\right)^2\,.
\end{align*}
Choosing $L=n-2$, and consequently $\mch_{L}$ to correspond to the first $n-1$ qubits with $d_L=2^{n-1}$, we arrive at $(d_L/d)^2 = {1}/{4}$. Therefore, we conclude that, in order to form a $2$-design, an ensemble of $2$ local gates over these groups must have depth $L> n-2$.  This proves Corollaries~\ref{crl:ortho} and~\ref{crl:symp}.

\subsection{Clifford group}~\label{app:clifford}
Next we turn to the important case of the Clifford group. Interestingly, and in contrast to the other groups we consider, we are here able to rule out sublinear-depth approximate designs only at $k\geq 8$. The core technical result that we will need is
\begin{lemma}
[Corollary of Lemma VII.1 in Ref.~\cite{montealegre2022duality}]\label{prop:cliff}
    There exist isometries $V_1,V_2$ such that $\overline{U}=V_1U^{\ot 7}V_2$.
\end{lemma}
Given Lemma~\ref{prop:cliff}, we can readily construct an experiment on eight copies of $\mch$ that distinguishes between ensembles over sublinear-depth Cliffords and Clifford 8-designs. Indeed, consider the following experiment. We assume we are given access to a sample $C^{\ot 8}$ where either $C\sim\mu_{\rm Cliff}$ or $C\sim \mce_{L}$, where $\mce_L$ is an ensemble of 2-qubit-generated depth-$L$ Cliffords. First, use Lemma~\ref{prop:cliff} to map the last seven copies of $C$ to $\overline{C}$; concretely, note that we can apply  $C\ot\overline{C} = C\ot(V_1 C^{\ot 7}V_2)$ which is a unitary map $\mch\ts\to\mch\ts$. Further noting that the Bell state $\ket\Phi$ over these two copies of $\mch$ (which we decompose as $\mch\ts=\mch_L\ot\mch_{\overline{L}}\ot\mch_L\ot\mch_{\overline{L}}$) is in this representation Clifford-invariant, apply $C\ot\overline{C}$ to $(V\ot\id)\ket\Phi$ for some local perturbation $V$, and measure the POVM $\{\Pi_{\mce_L}, \id-\Pi_{\mce_L}\}$, where 
\begin{equation}
    \Pi_{\mce_L} = \id_{\mch_L\ts}\ot\ketbra{\Phi_{\overline{L}}} ,\qquad \Pi_G = \id - \Pi_{\mce_L}.
\end{equation}
By the usual reasoning $p(\Pi_{\mce_L}\hspace{0.7mm}|\hspace{0.7mm}U\sim\mce_L)=1$, and 
\begin{align}
    p(\Pi_{\mce_L}\hspace{0.7mm}|\hspace{0.7mm}U\sim\mce_G) &=\Tr\left[\expect_{U\sim\mu_{\rm Cliff}}(U\ot U^*) Z_1\ketbra{\Phi}Z_1(U^\dagger\ot U^\mst) \ketbra{\Phi_{\overline{L}}}\right]\nonumber\\
&=\Tr_{\overline{L}}\left[\Tr_L\left[\expect_{U\sim\mu_\mbu}(UZ_1U^\dagger)\ketbra{\Phi}(UZ_1U^\dagger)\right]\ketbra{\Phi_{\overline{L}}}\right]\nonumber\\
&=\frac{1}{dd_{\overline{L}}}\Tr[\phi_{\mu_\mbu}^{(2)}\left(Z_1\ot Z_1\right)\mss_L]\nonumber\\
&\approx d_{\overline{L}}^{-2}\,,
\end{align}
where we have used that the Cliffords form a unitary 2-design to evaluate the average. Taking (say) $L=n-2$, we have $d_{\overline{L}}^{-2}=1/4$, and that therefore 
\setcounter{crl}{3}
\begin{crl}
Any ensemble built from $2$-local Clifford gates requires depth $L\geq n-1$ to form an approximate Clifford $8$-design.
\end{crl}

\subsection{Mixed-unitary}~\label{app:mu}
To finish, we consider the case of  mixed-unitary circuits, which we recall to consist of elements of the form $U\ot U^*$, where $U$ is an $n$-qubit unitary. 
We have an invariant state $\ket\Phi\in\mch$,  the Bell state across the first and second groupings of $n$ qubits, which is manifestly mixed-unitary-invariant. 
In a standard application of our procedure we take the perturbation to be $Z_1$, the Pauli-$Z$ operator on the first qubit, introduce the partition $\mch=\mch_L\ot\mch_{\overline{L}}\ot\mch_L\ot\mch_{\overline{L}}$, and let
\begin{equation}
    \Pi_{\mce_L} = \id_{\mch_L\ts}\ot\ketbra{\Phi_{\overline{L}}} ,\qquad \Pi_G = \id - \Pi_{\mce_L}.
\end{equation}
By the usual reasoning $p(\Pi_{\mce_L}\hspace{0.7mm}|\hspace{0.7mm}U\sim\mce_L)=1$, and (with $d=2^n$ and recalling the calculation of the previous subsection)
\begin{align}
    p(\Pi_{\mce_L}\hspace{0.7mm}|\hspace{0.7mm}U\sim\mce_G) &=\Tr\left[\expect_{U\sim\mu_{\mbu}}(U\ot U^*) Z_1\ketbra{\Phi}Z_1(U^\dagger\ot U^\mst) \ketbra{\Phi_{\overline{L}}}\right]\nonumber\\
&\approx d_{\overline{L}}^{-2}\,.
\end{align}
Hence, we find
\setcounter{crl}{4}
\begin{crl}
Any ensemble built from $4$-local mixed-unitary circuits (i.e., $U\ot U^*$ with $2$-local $U$) requires depth $L\geq n -1$ to form an approximate mixed-unitary $1$-design.
\end{crl}

In fact, this bound is optimal up to constant factors. Indeed, let $G$ be a group and $\HC$ a $G$-module via $R:G\arr \mbb{U}(\HC)$. For a subset $S\subseteq {\rm Im}(R)$ to form a (one)-design over such a group representation, we require $\LC(\HC)^G = \LC(\HC)^S$. Suppose $S$ forms a group. A natural way to determine whether $S$ is a design is to compare $\HC$ as a $G$-module and as a (restricted)$S$-module.
Recall that any $G$-module, $\HC$ has a decomposition into irreducible representations: 
\[
\HC\cong \bigoplus_\a V^\a_G \otimes \mbb{C}^{m_\a}\,.
\]
Similarly, $\HC \downarrow_S$ will decompose into $S$ irreps. Via Schur's lemma, since $\LC(\HC)=\HC\otimes \HC^*$, there is a linearly independent invariant in $\LC(\HC)^G$ for each pair of isomorphic irreps in $\LC$, whence   $\dim(\LC(\HC)^G) = \sum_\a m_\a^2$. Thus, the condition of $S$ being a design over $G$-module $\HC$ reduces to the twofold condition that (i) each $G$-irrep $V^\a_G$ in $\HC$, when restricted $(V^\a_G)\downarrow_S$, is irreducible as a $S$-module, and (ii) distinct $G$-irreps do not become isomorphic under the restriction.

We now turn specifically to mixed unitaries. Let $\HC=\mbb{C}^d\otimes \mbb{C}^d$ be a module for $\mbb{U}(d)$ via the group homomorphism $R(U) = U\otimes U^*$.
With $d=2^n$, 
let $\ket{\Phi} = d^{-1/2}\sum_{i=1}^d \ket{ii} \coloneqq \ket{I}$ be the Bell state, and denote $\ket{P}=(I\otimes P) \ket{I}$ for any $P\in \PC_n$ the set of $n$-qubit Paulis. Note that the set $\{\ket{P}\}_{P\in \PC_n}$ forms an orthonormal basis for $\HC$. Let ${\rm Cl}_n$ denote the $n$-qubit Clifford group, the normalizer of $\PC_n$. Consider the action of $\mc{P}_n$ and ${\rm Cl}_n$ on $\HC$ via restriction of $R$. We claim that mixed-Paulis fail to form mixed-unitary one-designs, while mixed-Clifford do.

\begin{prop}\label{prop:mu}
$\PC_n$ does not form a mixed-unitary one design, but ${\rm Cl}_n$ does.
\end{prop}

\begin{proof}
Note that as a $\mbb{U}(d)$-rep, $\HC\cong \HC_{\triv} \oplus \HC_{\rm adj}$, with two irreps: $\HC_{\rm triv}=\spn\{\ket{I}\}$ a trivial and $\HC_{\rm adj} =\spn\{\ket{P}\}_{P\in \PC_n\setminus\{I\}}$ an adjoint. The statement of the proposition follows from the fact that, although the trivial irrep is again an irrep (the trivial) when restricted to both  $\PC_n$ and ${\rm Cl}_n$ (and indeed to any group), the adjoint unitary irrep is a) irreducible when restricted to ${\rm Cl}_n$ (it is not hard to see that $\HC_{\rm adj}$ contains no proper invariant subspaces under ${\rm Cl}_n$), but b) completely shattered into $d^2-1$ one-dimensional irreducible representations when restricted to $\PC_n$:
\[
\HC \cong \bigoplus_{P\in \PC_n} \HC_P
\]
where for each $P$, $\HC_P =\spn \{\ket{P}\}$ is a one-dimensional irreducible representation -- $R(Q)\ket{P} =\ket{QPQ^{-1}} = \pm \ket{P}$ for all $Q\in \PC_n$.

Alternatively, we can note that from Eq.~\eqref{eq:mixedudecomp} the uniform ensemble on ${\rm Cl}_n\ot {\rm Cl}_n^*$ is a mixed-unitary one-design if and only if there are exactly two operators in $\mcl(\mch)$ that commute with $C\ot C^*$ for all $n$-qubit Cliffords $C$. Equivalently, ${\rm dim\ }\mcl(\mch)^{{\rm Cl}_n\ot {\rm Cl}_n^*}=2$. As this dimension is given by the trace of the (vectorized) first-order moment operator,
we can directly calculate
\begin{align}
{\rm dim\ }\mcl(\mch)^{{\rm Cl}_n\ot {\rm Cl}_n^*}&= \Tr[{\rm vec\ }\phi^{(1)}_{{\rm Cl}_n \ot {\rm Cl}_n^*}]\nonumber\\
&=\frac{1}{|{\rm Cl}_n|}\Tr[\sum_{C\in {\rm Cl}_n} (C\ot C^*)\ot(C\ot C^*)^*]\nonumber\\
&=\frac{1}{|{\rm Cl}_n|}\sum_{C\in {\rm Cl}_n} \lvert\Tr[C]\rvert^4\nonumber\\
&=2
\end{align}
where the final equality follows from the fact that the Cliffords are a unitary 2-design~\cite{mele2023introduction}. 
\end{proof}

A consequence of Proposition~\ref{prop:mu} together with Corollary~\ref{crl:mixedu} is that, up to constant factors, the mixed-Clifford group is an optimal mixed-unitary one-design.

\section{Further representation-theoretic details}\label{sec:rep}
In this section we review some relevant technical details of representation theory. Many more details can be found in, for example, Ref.~\cite{fulton1991representation}. 
We will primarily be occupied with the fact that irreducible representations admit the following classification into ``real'', ``complex'', and ``quaternionic'' representations~\cite{fulton1991representation}:
\begin{fact}\label{fact:crq}
An irreducible representation $\mcv$ is one (and only one) of the following:
\begin{enumerate}
    \item Complex: $\chi_\mcv$ is not real-valued; $\mcv$ does not have any $G$-invariant non-degenerate bilinear forms.
    \item Real: $\mcv = \mcv_0 \otimes_{\mbr} \mbc $, $\mcv_0$ a representation over $\mbr$; $\mcv$ has a $G$-invariant symmetric non-degenerate bilinear form.
    \item Quaternionic: $\chi_\mcv $ is real, but $\mcv$ is not; $\mcv$ has a $G$-invariant antisymmetric non-degenerate bilinear form.
\end{enumerate}
\end{fact}
Here $\chi_\mcv=\Tr \circ R_\mcv:G\to\mbc $ is the \textit{character} of $\mcv$, i.e., the trace of the representing elements $R_\mcv:G\to \mcl(\mcv)$. 
In fact, it is not too hard to see explicitly that for an irreducible representation $\mcv$, the trichotomy \{no preserved bilinear form, a symmetric preserved bilinear form, or an antisymmetric preserved bilinear form\} exhausts all the possibilities. That is,
\begin{lemma}
Let $\mcv$ be an irreducible unitary representation of a group $G$ with a non-degenerate preserved bilinear form $\Om$.  Then $\Om$ is either symmetric or antisymmetric. 
\end{lemma}
\begin{proof}
We give an elementary proof, inspired by a similar argument in Ref.~\cite{zee2016group}. First recall that from Lemma~\ref{lem:supp3} we can without loss of generality take $\Om$ to be unitary, and hence invertible.  Let $U_g\in\mbu(\mcv)$ denote the representing element of some $g\in G$. As is well-known, we have $U_{g^{-1}}=U_g^{-1}$; the presumed $G$-invariance of $\Om$ then yields $U_g^{-1} \Om (U_g^{-1})^\mst=\Om$, from which we deduce the  relations
  $ \Om^{\mst}U_g^*=U_g \Om^\mst$ and $\Om^{-1}U_g =U_g^*\Om^{-1}$, and that therefore
\begin{equation}
    \Om^\mst\Om^{-1} U_g = \Om^{\mst} U_g^* \Om^{-1} =U_g\Om^\mst\Om^{-1};
\end{equation}
i.e., $\Om^\mst\Om^{-1}$ commutes with  $U_g$ for every $g\in G$. So, by Schur's lemma, we have $\Om^\mst\Om^{-1} =\lm\id$ for some $\lm\in\mbc$; from $\Om=(\Om^\mst)^\mst = \lm^2\Om$ it further follows that $\lm=\pm 1$, i.e., $\Om$ is either symmetric or antisymmetric.
\end{proof}

For our purposes the most interesting aspect of the classification of Fact~\ref{fact:crq} is  the correspondence between these classes and the (non-)existence of a preserved bilinear form. Remarkably, there exists a simple numerical invariant associated to an irrep that identifies its class with respect to the above possibilities; indeed, the
\textit{Frobenius-Schur indicator} $\mcf(\mcv)$
is given by~\cite{fulton1991representation}
\begin{equation}\label{eq:crq}
\mcf(\mcv)=\int_{G}d\mu_G(g)\ \chi_\mcv(g^2) = \begin{cases}
    0&{\rm \ if\ } \mcv{\rm \ is\ complex}\,,\\
    1&{\rm \ if\ } \mcv{\rm \ is\ real}\,,\\
    -1&{\rm \ if\ } \mcv{\rm \ is\ quaternionic}\,.\\
\end{cases}
\end{equation}
Let us verify the behaviour of the Frobenius-Schur indicator on the standard representations of the unitary, orthogonal, and symplectic groups, equipped with their uniform measures. In all cases we can proceed mechanically by means of the Weingarten-calculus~\cite{mele2023introduction,collins2009some,collins2006integration} over the group under consideration; in the case of the unitary group, however, we have a particularly simple argument. By the left-invariance of the unitary Haar measure $\mu_{\mbu(d)}$, in particular under the transformation $U\mapsto e^{i\theta}U$ for all $\theta\in \mbr$, we have
\begin{equation}
\int_{\mbu(d)}d\mu_{\mbu(d)}(U)\ \Tr[U^2] = \int_{\mbu(d)}d\mu_{\mbu(d)}(U)\ \Tr[(e^{i\theta}U)^2] = e^{2i\theta}\int_{\mbu(d)}d\mu_{\mbu(d)}(U)\ \Tr[U^2];
\end{equation}
taking $\theta=\pi/2$, for example, we conclude that the integral must vanish, and  therefore by Eq.~\eqref{eq:crq} that the standard representation of the unitary group is \textit{complex}. Fact~\ref{fact:crq} then tells us that the unitary group does not possess an invariant bilinear form; of course, this is consistent with what we already know.

In the case of the orthogonal group we rely on the happy fact that one can obtain (by means of the aforementioned  Weingarten-calculus) explicit expressions for integrals of products of matrix elements over the orthogonal group. In particular, one has~\cite{collins2009some}
\begin{equation}\label{eq:ow}
\int_{\mbo(d)}d\mu_{\mbo(d)}(O)\  O_{ab} O_{ba} = \frac{\delta_{a,b}}{d}.
\end{equation}
Picking any basis, we then have
\begin{equation}
\int_{\mbo(d)}d\mu_{\mbo(d)}(O)\ \Tr[O^2] = \int_{\mbo(d)}d\mu_{\mbo(d)}(O)\ \sum_{a,b=1}^d O_{ab} O_{ba}= \sum_{a,b=1}^d \frac{\delta_{a,b}}{d} = 1,
\end{equation}
and Fact~\ref{fact:crq} then assures us that the standard representation of the orthogonal group is real. 
Finally,  directly carrying out the first order integral over the symplectic group yields
\begin{equation}
\int_{\mbsp(d/2)}d\mu_{\mbsp(d/2)}(S)\  S_{ab} S_{ba} = \frac{\Om_{ab}\Om_{ba}}{d},
\end{equation}
from which we quickly see that the standard representation of the  symplectic group is quaternionic.

More generally, it is not hard to see that the Frobenius-Schur indicator is \textit{additive} over irreducible components; that is, if $\mcv\cong\bigoplus_{\lm=1}^{\Lm} \mcv_\lm^{\oplus m_\lm}$ is a decomposition of $\mcv$ into mutually-inequivalent irreps $\mcv_\lm$, then we have
\begin{equation}\label{eq:fsred}
    \mcf(\mcv)= \sum_{\lm=1}^{\Lm}m_\lm\mcf(\mcv_\lm).
\end{equation}
Equation~\eqref{eq:fsred} allows us to reason about groups acting reducibly. For example, consider the representation $D$ of $\mbu(2^n)$ on a $2n$-qubit Hilbert space by $D:U\mapsto U \ot U^*$. We can calculate the number of irreps in $D$ by the well-known formula~\cite{fulton1991representation}
\begin{equation}\label{eq:mixedudecomp}
\sum_{\lm=1}^{\Lm^{(D)}}\left(m_\lm^{(D)}\right)^2 =\int_{\mbu(d)}d\mu_{\mbu(d)}(U)\ \lvert \chi_D(U) \rvert^2=  \int_{\mbu(d)}d\mu_{\mbu(d)}(U)\ \lvert \Tr[U\ot U^*] \rvert^2=\int_{\mbu(d)}d\mu_{\mbu(d)}(U)\  \Tr[U\ts\ot (U^*)\ts] = 2,
\end{equation}
where the final equality comes from another standard application of the Weingarten calculus. Evidently, the only possibility is that $D$ decomposes into the sum of two irreps, each occurring with multiplicity one. One of these irreps is easy to identify: the Bell state across the bipartition into the first and last $n$ qubits furnishes a trivial one-dimensional representation $D_0$ of $\mbu(d)$; the remaining irrep is then its orthogonal complement $(D_0)^\perp$ in $(\mbc^2)^{\ot 2n}$. The Frobenius-Schur indicator of $D$ is also easily calculated by Weingarten calculus (combined with the \textit{swap-trick}~\cite{mele2023introduction}):
\begin{equation}
    \mcf(D) = \int_{\mbu(d)}d\mu_{\mbu(d)}(U)\  \Tr[U^2\ot (U^*)^2] = 2.
\end{equation}
From the previous discussion we have that $\mcf(D)=\mcf(D_0)+\mcf((D_0)^\perp) = 1 + \mcf((D_0)^\perp)$. Hence, we conclude that $\mcf((D_0)^\perp) = 1 $, and that therefore $D$ decomposes into two real representations. Evidently, the existence of preserved bilinear forms on each of the  irreducible components of a representation implies the existence of a globally preserved bilinear form; the converse, however is false. To see this, let us consider the example of matchgate circuits,  generated as we recall by single qubit $Z$ rotations and nearest-neighbor $XX$ rotations~\cite{diaz2023showcasing}. Evidently the subspaces $\mch_{\rm even}$ ($\mch_{\rm odd}$) of  $\mch$ spanned by the computational basis states with even (odd) Hamming weight furnish representations of the matchgate group; in fact they are irreducible, corresponding to the \textit{spinor representations} of ${\rm Spin}\,(2n)$ with highest weights $(1/2,1/2,\ldots \pm 1/2)$~\cite{fulton1991representation}. Their behaviour (insofar as the above  discussion goes) turns out to depend exactly on $n\ {\rm mod\ } 4$; specifically, we have~\cite{fulton1991representation}
\begin{equation}
    \mch_{\rm even},\ \mch_{\rm odd} = \begin{cases}
        {\rm real}\qquad & n\equiv 0\ \hspace{4.5mm} ({\rm mod\ } 4)\,,\\
        {\rm complex}\qquad & n\equiv 1,3\ \ ({\rm mod\ } 4)\,,\\
        {\rm quaternionic}\qquad & n\equiv 2\hspace{5.6mm} ({\rm mod\ } 4)\,.
    \end{cases}
\end{equation}
But, as discussed above, the matchgate group possesses a preserved bilinear form for all $n$. This presents no difficulty to our above claims, however; indeed it is not difficult to see that when $n$ is odd the bilinear form maps \textit{between} $\mch_{\rm even}$ and $\mch_{\rm odd}$, and is the zero map when restricted to one or the other. Thus (continuing to take $n$ to be odd), as expected, there is no (non-trivial) preserved bilinear form on either of those complex irreducible representations; being mutually dual~\cite{fulton1991representation}, however, their direct sum $\mch$ is self-dual, and does possess a so-preserved bilinear form (as indeed we saw in Appendix~\ref{sec:forms}).

\end{document}